\documentstyle[pre,aps,preprint,epsfig]{revtex}

\global\firstfigfalse
\tighten

\begin{document}

\draft
\title{Directed polymers in high dimensions}
\author{Ralf Bundschuh\footnotemark\ 
and Michael L{\"a}ssig}
\address{
Max-Planck-Institut f\"ur Kolloid- und Grenzfl\"achenforschung,
Kantstr.~55, 14513 Teltow, Germany}
\date{\today }
\maketitle
\footnotetext[1]{electronic address:
{\tt Bundschuh@mpikg-teltow.mpg.de}}

\begin{abstract}
We study directed polymers subject to
a quenched random potential in $d$ transversal dimensions.
This system is closely related to the Kardar--Parisi--Zhang
equation of nonlinear stochastic
growth.
By a careful analysis of the perturbation theory we show that
physical quantities develop singular behavior for $d\to4$. For
example, the universal finite size amplitude of the free energy at
the roughening transition is proportional to $\sqrt{4-d}$. This shows
that the dimension $d=4$ plays a special role for this system and
points towards $d=4$ as the upper critical dimension of the
Kardar--Parisi--Zhang problem.
\end{abstract}
\pacs{PACS number(s): 05.40+j, 64.60.Ht, 05.70.Ln}

\narrowtext

\section{Introduction}

The field of non-equilibrium growth processes has attracted quite a
lot of interest in the recent past~\cite{krug92}.
The simplest nonlinear model describes a growing
surface on sufficiently large length scales as a height profile
$h(r,t)$ over a $d$--dimensional reference plane parameterized by $r$.
The dynamics of this surface is given by the
Kardar--Parisi--Zhang~\cite{kar86} (KPZ) equation
\begin{equation}\label{kpz}
{\partial\over\partial t} h =\nu\Delta h+\frac12\lambda(\nabla h)^2
+\eta
\end{equation}
with a Gaussian white noise defined by
\begin{equation}\label{discor}
\overline{\eta(r,t)}=0 \mbox{ and }
\overline{\eta(r,t)\eta(r^\prime,t^\prime)}=
\sigma^2\delta^d(r-r^\prime)\delta(t-t^\prime).
\end{equation}
The surface described by~(\ref{kpz}) can be in different phases which
depend on the dimension-less coupling constant
\begin{equation}\label{gdef}
g=-{\lambda^2\sigma^2\over2\nu^3}.
\end{equation}
and the space dimension $d$. In less than two dimensions there are
two different phases: the weak coupling phase for $g=0$ and the strong
coupling phase for $g\not=0$. Above two dimensions, there exists a
critical value $g_c$. The surface is in the weak coupling phase for
$|g|<g_c$ and in the strong coupling phase for $|g|>g_c$. Precisely
at $|g|=g_c$ the system undergoes a roughening transition.
Whereas the linear growth equation for $g=0$ can             
easily be solved in any dimension, it is much more difficult to get
information on the critical behavior of the other phases.

The morphology of the surface in the different phases is
characterized by the  asymptotic scaling of the
height-correlation-function
\[
\langle(h(r_1,t_1)-h(r_2,t_2))^2\rangle\sim
|r_1-r_2|^{2\chi}f(t|r_1-r_2|^{-z}),
\]
which defines the roughening exponent $\chi$ and the dynamic exponent
$z$. For $g\not=0$ these exponents satisfy the
hyperscaling relation $z+\chi=2$~\cite{med89}.

In the strong coupling phase it is known that
$z=3/2$ in $d=1$~\cite{fors77} and $z=2$ in the limit of infinite
dimension~\cite{derr88}.
In general dimensions all exact methods fail
and only numerical and mode--coupling results
are available, but they become less reliable in higher dimensions.
Therefore there is still a very controversial discussion
\cite{cook89,feig89,halp90,moor95,frey95}
about the existence of a finite upper critical
dimension of the KPZ problem, i.e. a dimension above which the
dynamical exponent $z$ has the constant value $2$.

In the language of the renormalization group, the different phases
belong to different fixed points. For less than two dimensions,
there is one unstable fixed point at $g=0$, which governs the
weak coupling phase, and one stable fixed point at $g\to-\infty$,
which governs the strong coupling phase. Above two dimensions,
the weak coupling fixed point also becomes stable and a new
fixed point describing the roughening transition appears in between
the other fixed points.

It has been shown~\cite{laes95} that the strong--coupling fixed point
is inaccessible by a perturbation expansion around $g=0$.
The situation is somewhat better for the fixed point describing the
roughening transition. The singularities, which arise in the
perturbation series above two dimensions, can be treated in a
systematic expansion~\cite{laes94a,raja91} with parameter
\begin{equation}
\epsilon={2-d\over2}.
\end{equation}
In the framework of this $\epsilon$-expansion,
one finds the exponents $z^*=2$ and $\chi^*=0$, which are exact to all
orders in perturbation theory~\cite{laes95}. Hence,
hyperscaling is preserved.

However, this perturbation expansion breaks down for $d\to4$ since new
singularities in the coefficients of the series expansions arise at
$\epsilon^\prime=0$, where
\begin{equation}
\epsilon^\prime\equiv\epsilon+1={4-d\over2}.
\end{equation}
The treatment of these singularities is the aim of this
paper. We develop a systematic way to extract the behavior of physical
quantities as $d\to4$ from the divergent series in $\epsilon^\prime$.
We show that in contrast to the singularities in $\epsilon$, these
singularities translate into a non-analytic behavior of observable
quantities. This stresses the importance of $d=4$ for the KPZ
problem and therefore shows that four
dimensions is a good candidate for an upper critical dimension.

We will address this question not in the KPZ picture but by using the
exact Hopf--Cole--mapping~\cite{hopf50,cole51} of the KPZ problem
to a directed polymer
in a random medium.
A directed polymer (also called a string) is a line with a
preferred direction, which is governed by its line tension. The
energy of a given conformation is
\begin{equation}\label{randhamil}
{\cal H}[r]=\int_0^{L_{||}}\frac12\dot r^2(t)
+\lambda\nu^{-3/2}\eta(r(t),t){\rm d}t,
\end{equation}
where $r(t)$ is a $d$-dimensional vector which
denotes the elongation of the directed polymer at a position
along the preferred (``time'') axis $t$, $L_{||}$ is the
projected length of the directed polymer
and $\eta$ is the random potential which appears in the KPZ equation.

To study the structure of the new singularities in the perturbation
series, we will use an especially simple physical quantity.
We nevertheless expect,
that more complicated quantities such as correlation functions
show the same type of singularities. We will study a system where
the projected length $L_{||}$ of the directed polymer is infinite
while its transversal ``motion'' is restricted to a finite volume of
a characteristic width $L_\perp\equiv L^{1/2}$. Within this system we
will concentrate on the dimension-less averaged free energy per unit
``time'', which we call
\begin{equation}\label{randomc}
{\cal C}(g,L)=
\lim_{L_{||}\to\infty}{L\over L_{||}}\overline{F(g,L,L_{||})}.
\end{equation}
The infrared regularization by the length scale $L$ has to be
introduced, because the series expansion starts at the Gaussian fixed
point ($g=0$), which has no intrinsic length scale. $L$ moreover
serves as the flow parameter of the renormalization group considered
below.

Since at the roughening transition hyperscaling is preserved at least
around $2$ dimensions~\cite{laes95},
therefore there are no corrections to the $1/L_\perp^2=1/L$
behavior of the free energy per length
$\overline{F(g,L,L_{||})}/L_{||}$, and the finite size amplitude
\begin{equation}
{\cal C}(g)=\lim_{L\to\infty}{\cal C}(g,L)
\end{equation}
is a universal quantity.
Near $d=2$, one finds~\cite{laes95} for its value ${\cal C}^*$ at the
unbinding transition 
\begin{equation}
{\cal C}^*(\epsilon)\sim\epsilon+O(\epsilon^2).
\end{equation}

$\cal C$ is not only one of the simplest physical quantities to be
calculated in our system, but also one of the most fundamental ones:
it plays a role very similar to the central charge
in two-dimensional conformal models~\cite{card86}.

The directed polymer problem with randomness described
by~(\ref{randhamil}) can be treated analytically via the replica
trick. This means that it can be expressed
as the limiting case $N\to0$ of $N$ directed polymers
with a short-ranged interaction potential (see e.g.~\cite{kar87a}):
\[
{\cal H}[r_1,\ldots,r_N]=\int_0^{L_{||}}
\frac12\sum_{i=1}^N\dot r_i^2(t)+
g\sum_{i<j}^N\delta(r_i(t)-r_j(t)){\rm d}t.
\]
By~(\ref{gdef}), the potential is always attractive~($g<0$) and
captures the short--ranged correlations of the
disorder~(\ref{discor}).

The finite size amplitude~(\ref{randomc}) of the free energy per unit
``time'' of the random system can then be obtained from
the limit $N\to0$ of the finite size
amplitudes per unit ``time'' and per pair of directed polymers
\begin{eqnarray*}
{\cal C}(g,L,N)&=&{2\over N(N-1)}\\
&&\times\!\lim_{L_{||}\to\infty}
{L\over L_{||}}[F(g,L,N,L_{||})-\!F(0,L,N,L_{||})]
\end{eqnarray*}
of the $N$-polymer system without randomness.

The development of a new regularization scheme for the perturbation
theory of $\cal C$ near 4 dimensions is done in two steps. In the
first step, the fact is used that for only two directed polymers
independent
techniques exist, that allow an exact solution of the
2--polymer--problem.
Those methods (the transfer matrix approach and the resummation
of the perturbation series) are reviewed and extended to
our system with a transversally restricted ``movement''
in section~\ref{twopol} of this paper

In order to be able to get results about perturbation series where
no exact solution exists any more, a regularization scheme is
developed by comparison with the exact solution. The regularization
scheme itself then is independent of the existence of the exact
solution in the sense that it extracts the results directly from the
coefficients of the new singularities in the perturbation series.
It is shown that only the most divergent
terms in every order of perturbation theory contribute to the first
order terms of the limiting $\beta$ function.

As the second step this regularization scheme is then applied to the
case of an arbitrary number of directed polymers
in section~\ref{arbitN}, which can only be solved
exactly in one dimension~\cite{kar87}. A diagrammatic expansion of the
partition function and of the free energy is developed.
The series expansion is calculated up to the fifth order in the
coupling constant and many simplifications appear, that suggest a
simple underlying structure.

Since the most important diagrams with respect to the regularization
scheme turn out to be just 2-polymer-diagrams, the main
properties of the $\beta$ function can still be extracted from
the general structure of the perturbation series.

Independent of the number of directed polymers in
less then 4 dimensions there are two zeroes of the $\beta$ function:
one is the finite size amplitude ${\cal C}^*$ of the free energy per
unit ``time'' at the unbinding transition; the other one is
at the Gaussian fixed point.
In 4 dimensions the $\beta$ function becomes just a straight line
with a fixed slope. ${\cal C}^*$ and the zero at the Gaussian
fixed point coincide there. We find that near 4 dimensions
the physical quantity ${\cal C}^*$ behaves non--analytically as
\begin{equation}\label{zeroapp}
{\cal C}^*(\epsilon^\prime) \sim \sqrt{\epsilon^\prime}.
\end{equation}

Since the behavior is independent of the number of directed polymers,
it should
also describe the scenario in the replica limit of a vanishing number
of directed polymers and therefore the behavior of the weak-coupling
fixed point of the KPZ problem.

In section~\ref{conclusion}, the main results are summarized again.
Many of the technical details have been postponed to various
appendices to keep the route of argumentation straight.

\section{Two directed polymers}\label{twopol}

Since the problem of two directed polymers ($N=2$)
is especially simple, we start
examining this problem and try to generalize our results in the next
section.

The reason for the simplicity of the two polymer problem is that
one can separate the problem in the motion of a ``center of mass''
$({\bf r}_1(t)+{\bf r}_2(t))/2$, which just fluctuates freely
independent of the coupling constant $g$ and the problem in the
relative coordinate $\hat r\equiv{\bf r}_2-{\bf r}_1$. Since the
KPZ--problem is equivalent to
directed polymers with a relative short--range interaction
($V(r_1,r_2)=g\delta^d(r_1-r_2)$),
we have to deal in the relative coordinate with one directed polymer,
which is interacting with a straight line defect
at the origin of the perpendicular space. The transition
probability from ${\bf r}^\prime$ to
${\bf r}$ within the projected contour
length $t$ (the restricted partition function) is given by the
single path integral
\begin{equation}\label{zpathint}
Z_t({\bf r}^\prime,{\bf r})=\int_{(0,{\bf r}^\prime)}^{(t,{\bf r})}
{\cal D}\hat r \exp\left[-\int_0^t
\frac12\dot{\hat r}^2(t^\prime)+
V(\hat r(t^\prime)){\rm d}t^\prime\right].
\end{equation}

The problem of one directed polymer interacting with a defect line
can now be treated in different manners: On the one hand the path
integral formulation of the directed polymer
problem is equivalent to a ``quantum mechanics'' and can
be solved in analogy to the corresponding quantum mechanical
problem. This is called the transfer matrix approach because the
Hamiltonian of the quantum mechanics is the continuum transfer matrix
of the problem.

The second approach just sets up a perturbative expansion of the
problem with the defect with the strength of the defect as the
expansion parameter, which can either be calculated coefficient
by coefficient or in the special case of two directed polymers can
even be summed up completely.

A comparison of these approaches shows that they lead to the
same results. Because we know the exact solutions, we can use them
to develop a treatment of the perturbation series in the interesting
case $\epsilon^\prime\to0$, which we afterwards can apply to problems
where no exact solution is available any more, as for example the
case of more than two directed polymers.

\subsection{Transfer matrix results}

Since the path integral~(\ref{zpathint}) is formally
a quantum mechanical path integral in imaginary time, the partition
function obeys the Schroedinger equation
\begin{equation}
(-\Delta+V({\bf r}))Z_t({\bf r}^\prime,{\bf r})=
-{\partial\over\partial t}Z_t({\bf r}^\prime,{\bf r}),
\end{equation}
of a particle in a potential $V({\bf r})$, which can be attacked by
standard quantum mechanical methods.

If the movement is restricted to a finite volume characterized by
some length $L_\perp=L^{\frac12}$, the long--time (large projected
length of the directed polymer)
behavior of the partition function is an exponential decay,
the decay rate of which is given by the ground state energy of the
quantum mechanical problem
\begin{equation}
(-\Delta+V({\bf r}))\psi({\bf r})=E_0\psi({\bf r}),
\end{equation}
The free energy per unit ``time'' for long polymers is therefore given
by the ground state energy plus exponentially small corrections.
The finite size coefficient can be extracted by studying the behavior
of the free energy for large system sizes $L_\perp$. At the Gaussian
fixed point we expect from dimensional analysis
that $E_0\sim\frac1{L_\perp^2}=\frac1L$.
The prefactor of this decay is ${\cal C}$.
In order to calculate ${\cal C}$, we extend here the calculations
in~\cite{lipo91} to the case of ``movement'' in a finite volume.

\subsubsection{Precise definition of the model}

In order to solve the model in arbitrary dimensions, we have to use
potentials and boundary conditions, which have a spherical symmetry.
In this case, we can separate the wave function in a radial and an
angular part. As it is well known from quantum mechanics, it is
convenient to define the radial function $\Phi(r)$ to be
$r^{(d-1)/2}$ times the radial part of the
total wave function. This function obeys for the lowest (zero) angular
momentum - as it should be the case for the ground state - the
radial Schroedinger equation
\begin{equation}\label{radsgl}
-{\partial^2\over\partial r^2}\Phi(r)+\!
{(d-3)(d-1)\over4}{1\over r^2}\Phi(r)
+\!V(r)\Phi(r)\!=\!\tilde E_0\Phi(r).
\end{equation}

We implement the attractive potential at the origin as
a well potential with a small but finite extension $a$:
\begin{equation}\label{radpot}
V(r)=\left\{
\begin{array}{rl}-\tilde V_0&0\le r<a\\0&a\le r\end{array}
\right.
\end{equation}
By defining $\phi(r)\equiv\Phi(ra)$,
$E_0\equiv a^2\tilde E_0$, $\bar V(r)\equiv a^2V(ra)$
and $V_0\equiv a^2\tilde V_0$, the ultraviolet cutoff $a$ can be
eliminated and everything is written in dimension-less variables.

The boundary conditions imposed to the directed polymer have to be
radially symmetric, too. The two possibilities studied here in detail
are on the one hand the Dirichlet boundary conditions, where
the directed polymer is put into a spherical box of 
radius $L_\perp$ with hard walls, which means $\phi(L_\perp/a)=0$,
and on the other hand the
von-Neumann boundary conditions, which
impose that the first derivative of the radial part of
the wave function vanishes at the boundary of a spherical box of
radius $L_\perp$, in order to mimic a kind of periodic boundary
conditions. This is equivalent to the condition
\begin{equation}
{L\over a}\left.{\partial\over\partial y}\right|_{y={\frac{L_\perp}a}}
\phi(y)={d-1\over2}\phi(\frac{L_\perp}a)
\end{equation}
for the radial function $\phi$.

Moreover we will consider harmonic boundary conditions. That means
that we allow a ``movement'' in infinite space, but add an harmonic
well potential $(\gamma/2)r^2$, which effectively restricts the
motion. The associated width of the box is from dimensional analysis
$\sqrt\gamma=(\pi/2)(1/L_\perp^2)$. (The factor of $\pi/2$
has only been introduced for convenience).

\subsubsection{Results for the free energy}

For the potential as given by~\ref{radpot}
and the first two boundary conditions given in the above section,
it is relatively easy, to calculate the ground state energy, because
the potential is piecewise constant. Only standard techniques of
quantum mechanics are applied. A sketch of these calculations is
given in appendix~\ref{appschroedsol}.

Here, we just give the results for the ground state energy.
Let us first consider the Dirichlet boundary conditions.

For the free system ($V_0=0$) the ground state
energy is exactly given by
\begin{equation}
E_0=x_{|\epsilon|}^2\left(\frac a{L_\perp}\right)^2,
\end{equation}
with $x_{|\epsilon|}$ being the smallest positive root of the
Bessel function $J_{|\epsilon|}$.
This is also the
asymptotic behavior for $0<V_0<V_*$. At the phase transition point
($V_0=V_*$) we have
\begin{equation}
E_0\approx\left\{\begin{array}{ll}
y_{|\epsilon|}^2\left(\frac a{L_\perp}\right)^2&|\epsilon|<1\\
4(|\epsilon|-1)\left(\frac a{L_\perp}\right)^{2|\epsilon|}&
|\epsilon|>1
\end{array}\right.
\end{equation}
where $y_{|\epsilon|}$ is the smallest positive root of
$J_{-|\epsilon|}$.

For von-Neumann boundary conditions the ground state energy is zero
at $V_0=0$. At $0<V_0<V_*$ the asymptotic behavior is
\begin{equation}
E_0\approx -{4|\epsilon|(|\epsilon|+1)\over
2|\epsilon|{J_{|\epsilon|}(\sqrt{V_0})\over
\sqrt{V_0}J_{|\epsilon|+1}(\sqrt{V_0})}-1}
\left(\frac a{L_\perp}\right)^{2(|\epsilon|+1)}.
\end{equation}
The decay is faster than quadratic which
means that the coefficient we are looking for remains zero.
This scaling behavior of the free energy per length in a system of
finite transversal size at the unbinding transition
is consistent with the scaling behavior of the eigenenergy of the
bound state in the infinite system~\cite{lipo91a}.

At the phase transition we get
\begin{equation}
E_0\approx\left\{\begin{array}{ll}-z_{|\epsilon|}^2
\left(\frac a{L_\perp}\right)^2&
|\epsilon|<1\\
4\sqrt{|\epsilon|(|\epsilon|^2-1)}
\left(\frac a{L_\perp}\right)^{|\epsilon|+1}&|\epsilon|>1
\end{array}\right.
\end{equation}
where $z_{|\epsilon|}$ is the smallest positive root of
$I_{-|\epsilon|-1}$. The decay for
$|\epsilon|>1$ is again faster than quadratic.

\begin{figure}[ht]
\begin{center}
\epsfig{figure=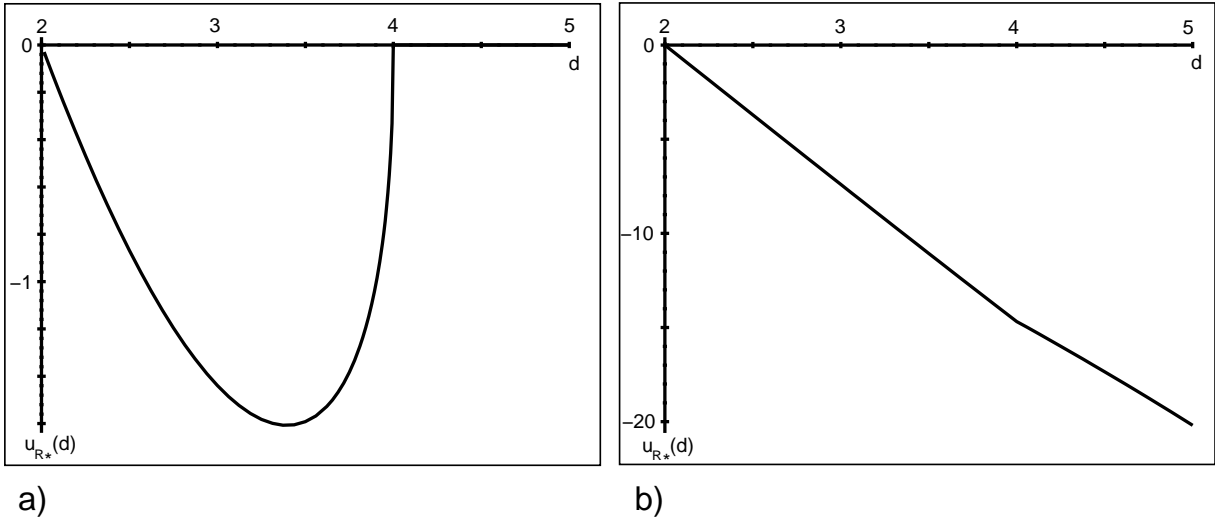,width=\textwidth}
\caption{Transfer matrix results for the renormalized coupling
constant. The dependence on the dimensionality is shown for
von-Neumann (a) and Dirichlet (b) boundary
conditions.\label{trmurplot}}
\end{center}
\end{figure}

The differences between the decay coefficients at the phase
transition point and in the free case (which we defined to be
${\cal C}^*$ in the introduction)
are plotted against the dimension in Fig.~\ref{trmurplot}.

It can be seen that ${\cal C}^*$ stays finite as $d$ approaches 4
with Dirichlet boundary conditions, whereas for
von-Neumann boundary conditions ${\cal C}^*$ approaches
zero. From the first three terms of the series expansion of the
Bessel function $I_{-|\epsilon|-1}$ (the third term is the first one
with a non-vanishing coefficient in the limit $|\epsilon|\to1$) we get
near $|\epsilon|=1$ a behavior of
$z_{|\epsilon|}\sim
4\sqrt{(1+|\epsilon|)(1-|\epsilon|)}+4(|\epsilon|-1)$,
which yields equation~(\ref{zeroapp}).

For harmonic boundary conditions the differential equation in the
outer area is a bit more complicated. Its solution is
\begin{equation}
\Phi(y)\sim y^{-\frac12}
W_{{E_0\over 4\sqrt\gamma a^2},\frac\epsilon2}(\sqrt\gamma a^2 y^2)
\end{equation}
with the Whittaker function $W_{\lambda,\mu}(z)$. The condition of
differentiability at $y=1$ is also more complicated to evaluate than
for the von-Neumann or Dirichlet boundary conditions, but in the end
we get
\begin{equation}
E_0\sim \pi\frac d2 \left(a\over L_\perp\right)^2
\end{equation}
at $0<V_0<V_*$ and
\begin{equation}
E_0\sim \pi\left(2-\frac d2\right) \left(a\over L_\perp\right)^2
\end{equation}
at the phase transition $V_0=V_*$. Therefore
\begin{equation}
{\cal C}=\pi(2-d)=2\pi\epsilon.
\end{equation}

\subsection{General perturbation theory}\label{sectdiag2}

Starting from the partition
\begin{equation}
{\cal H}={\cal H}_0+g\int_0^{L_{||}}\Phi(t){\rm d}t\quad\mbox{with}
\quad\Phi(t)\equiv\delta^d(r(t))
\end{equation}
into the free part (kinetic part and infrared regularization) and
an interaction part we can easily set up a perturbation series of
the partition function as
\begin{eqnarray}\label{Zseries}
{Z\over Z_0}&=&\langle e^{-g\int_0^{L_{||}}\Phi(t){\rm d}t}\rangle_0
\\\nonumber
&=&1+\sum_{n=1}^\infty(-g)^n
\int\limits_{
\rlap{\hbox{$\hspace{7mm}\scriptstyle
0\le t_1\le\ldots\le t_n\le L_{||}$}}}
{\rm d}t_1\ldots\int{\rm d}t_n\langle\Phi(t_1)\ldots\Phi(t_n)\rangle_0
\end{eqnarray}
where the expectation values $\langle\ldots\rangle_0$ are taken with
respect to ${\cal H}_0$ and $Z_0$ is the partition function just for
${\cal H}_0$. Here the factor of $n!$ has been canceled out by
introducing a time ordering.

The free energy is the negative logarithm of the partition function
and can also be expressed as a series in $g$ by systematically
expanding the logarithm. Since all correlation functions are
translationally invariant in the limit of large $L_{||}$, one of the
integrations just gives a factor of $L_{||}$ which cancels out since
we are interested in the free energy per length. As a result of this
series expansion we get to the first orders
\begin{eqnarray*}
\hbox to 7mm{\rlap{$\displaystyle
\lim\limits_{L_{||}\to\infty}{F(g)-F(0)\over L_{||}}=$}\hfill}&&\\
&&g\langle\Phi(0)\rangle_0-
g^2\int_0^\infty
\langle\Phi(0)\Phi(t)\rangle_0-\langle\Phi(0)\rangle_0^2
{\rm d}t+O(g^3).
\end{eqnarray*}
Since the structure of the singularities in the case $d\to4$ does not
show up in the second order, we have to extend the perturbation series
to higher orders. (Higher terms of this formula can be found in
appendix~\ref{appexpansions}.)

For two directed polymers with a $\delta$ interaction it is
now particularly simple
to calculate the time ordered correlation functions, because they can
be reduced to two-point-functions. Consider the propagator
$G_t(r_1,r_2)$ of the relative coordinate of the free (only infrared
regularized) directed polymers.
With this, the multi-point-function is expressed as
\widetext
\begin{equation}\label{mpointbygs}
\langle\Phi(t_1)\ldots\Phi(t_n)\rangle_0=
{\int{\rm d}r_1\int{\rm d}r_2 G_{t_1}(r_1,0)G_{t_2-t_1}(0,0)\ldots
G_{t_n-t_{n-1}}(0,0)G_{L_{||}-t_n}(0,r_2)\over
\int{\rm d}r_1\int{\rm d}r_2 G_{L_{||}}(r_1,r_2)}.
\end{equation}
\narrowtext
For long chain length we can drop end effects and assume $t_1$ and
$L_{||}-t_n$ to be large enough so that we can combine those terms to
a one point function
\begin{equation}
\langle\Phi(t_1)\ldots\Phi(t_n)\rangle_0\!
=G_{t_2-t_1}(0,0)\ldots G_{t_n-t_{n-1}}(0,0)\langle\Phi(0)\rangle_0
\end{equation}
If we here set $n=2$ we get the relation between the return
probability $G_t(0,0)$ and the two point function, so that we can
express the multi-point function by two point functions as
\begin{eqnarray*}
\langle\Phi(t_1)\ldots\Phi(t_n)\rangle_0&=&
{1\over\langle\Phi(0)\rangle_0^{n-2}}\\
&&\times\langle\Phi(t_1)\Phi(t_2)\rangle_0
\ldots\langle\Phi(t_{n-1})\Phi(t_n)\rangle_0.
\end{eqnarray*}

Using this formula, we can express all the coefficients of the
perturbation series as multiple integrals over products of two-point
functions. It is even more convenient to introduce the dimension-less
connected two point function as
\[
g_1(s)\equiv
{\langle\Phi(0)\Phi(Ls)\rangle_{0,c}\over\langle\Phi(0)\rangle_0^2}
\equiv{\langle\Phi(0)\Phi(Ls)\rangle_0-\langle\Phi(0)\rangle_0^2\over
\langle\Phi(0)\rangle_0^2}.
\]
If we now introduce a dimension-less coupling constant
\begin{equation}
u_0\equiv g L^\epsilon=g L\langle\Phi(0)\rangle_0,
\end{equation}
the finite size coefficient of the free energy per unit ``time'' reads
\begin{eqnarray}\label{urseries}
{\cal C}(u_0)&=&L\lim_{L_{||}\to\infty}
{F(g)-F(0)\over L_{||}}=\\\nonumber
&=&u_0-u_0^2\int_0^\infty g_1(s){\rm d}s\\\nonumber
&&+u_0^3\int_0^\infty\!\int_0^\infty\!\Big[
g_1(s_1)g_1(s_2)-g_1(s_1+s_2)\Big]{\rm d}s_1{\rm d}s_2\\\nonumber
&&+O(u_0^4),
\end{eqnarray}
where the time-ordered integration has been broken up into integrals
over the respective time differences.

As we can see, the finite size coefficient is equal to the
dimension-less coupling constant $u_0$ to first order. Singularities
in this series can only arise in the integrals of the higher order
terms. The renormalized coupling
constant must be defined such that all singularities in the
higher order terms are canceled. This is done by imposing the
renormalization point condition for the renormalized coupling constant
\begin{equation}
u_R(u_0)={\cal C}(u_0).
\end{equation}
This renormalization point condition has the advantage, that it
gives a physical meaning to the renormalized coupling constant.

Now we have to study the perturbation series~(\ref{urseries}) in
detail.
Already in the third order the essential difference between the
regularization scheme at $\epsilon=0$ and at $\epsilon^\prime=0$
becomes obvious. Near $\epsilon=0$, the second term in the third order
coefficient does not diverge at all, so that only the first term,
which obviously factorizes to the square of the second order
coefficient, produces divergences. This statement is true to all
orders of perturbation theory, which produces just a geometric series
of divergences. The especially simple structure of those divergences
guarantees that the $\beta$ function calculated to the second order
is exact to all orders of perturbation theory~\cite{laes95,laes94}.

Near $\epsilon^\prime=0$ the situation is totally different, because
here also the second term diverges. So in every order combinations of
different types of divergences occur, which do not factorize any more.

Since the terms in the perturbation series are becoming quite nasty
in higher orders, it is
convenient to introduce a graphical representation. Therefore we will
draw in the $n$-th order $n$ points in a line which represent the $n$
time stamps involved. For each function $g_1$ we then draw a line
connecting the two time stamps which are the arguments of the two
point function. With this representation we can sketch our series
expansion up to higher orders as it is done in
appendix~\ref{appexpansions}.

Careful inspection shows that all diagrams appear that
have at least one line passing each interval between two points and
do not have two lines leaving one point in the same direction.
The corresponding prefactors up to the seventh order are given by
\begin{equation}
(-1)^{\#P-l-1}\prod_{p\in P}l_p!,
\end{equation}
where $l$ is the number of lines in the diagram, $P$ is the set of
points in the diagram and for each $p\in P$ $l_p$ denotes the number
of lines which pass the given point.

This perturbation series can be further simplified by applying some
relations among the integrals over two point functions as for example
\begin{center}
\setlength{\unitlength}{0.008in}%
\begin{picture}(348,22)(36,789)
\thicklines
\put(380,800){\circle*{8}}
\put(180,800){\circle*{8}}
\put(200,800){\circle*{8}}
\put(220,800){\circle*{8}}
\put(360,800){\circle*{8}}
\put(275,795){\makebox(0,0)[lb]{\smash{=}}}
\put(135,795){\makebox(0,0)[lb]{\smash{+}}}
\put(320,800){\circle*{8}}
\put(340,800){\circle*{8}}
\put(240,800){\circle*{8}}
\put(220,800){\oval( 40, 20)[bl]}
\put(220,800){\oval( 40, 20)[br]}
\put(200,800){\oval( 40, 20)[tr]}
\put(200,800){\oval( 40, 20)[tl]}
\put( 70,800){\oval( 20, 10)[bl]}
\put( 70,800){\oval( 20, 10)[br]}
\put( 70,800){\oval( 60, 20)[tr]}
\put( 70,800){\oval( 60, 20)[tl]}
\put(330,800){\oval( 20, 10)[tr]}
\put(330,800){\oval( 20, 10)[tl]}
\put( 40,800){\circle*{8}}
\put( 60,800){\circle*{8}}
\put( 80,800){\circle*{8}}
\put(100,800){\circle*{8}}
\put(360,800){\oval( 40, 20)[tr]}
\put(360,800){\oval( 40, 20)[tl]}
\end{picture}
\end{center}
Those relations can be produced by a general mechanism via Laplace
transformation as explained in detail in appendix~\ref{appdiagrels}.

Applying them consequently enables us to get rid of all ``nested''
diagrams.
Since we are finally interested in the inverse series to this one,
we can just invert it by standard techniques order by
order and end up with the series expansion of $u_0(u_R)$
\begin{center}
\setlength{\unitlength}{0.008in}%
\begin{picture}(398,136)(-38,696)
\thicklines
\put(200,740){\circle*{8}}
\put(180,740){\circle*{8}}
\put(160,740){\circle*{8}}
\put(120,700){\circle*{8}}
\put(220,740){\circle*{8}}
\put(220,700){\circle*{8}}
\put(200,700){\circle*{8}}
\put(160,700){\circle*{8}}
\put(240,740){\circle*{8}}
\put( 40,700){\circle*{8}}
\put( 80,740){\circle*{8}}
\put( 60,740){\circle*{8}}
\put( 40,740){\circle*{8}}
\put(300,780){\circle*{8}}
\put(100,740){\circle*{8}}
\put( 60,700){\circle*{8}}
\put( 80,700){\circle*{8}}
\put(100,700){\circle*{8}}
\put(120,740){\circle*{8}}
\put(240,700){\circle*{8}}
\put(-38,816){\makebox(0,0)[lb]{\smash{$u_0(u_R) =$}}}
\put(214,776){\makebox(0,0)[lb]{\smash{+}}}
\put(110,776){\makebox(0,0)[lb]{\smash{+2}}}
\put(194,816){\makebox(0,0)[lb]{\smash{+}}}
\put( 55,816){\makebox(0,0)[lb]{\smash{+}}}
\put(130,736){\makebox(0,0)[lb]{\smash{+3}}}
\put( 16,696){\makebox(0,0)[lb]{\smash{+}}}
\put(135,696){\makebox(0,0)[lb]{\smash{+}}}
\put(251,736){\makebox(0,0)[lb]{\smash{+2}}}
\put(254,697){\makebox(0,0)[lb]{\smash{+}}}
\put(300,740){\circle*{8}}
\put(320,740){\circle*{8}}
\put(340,740){\circle*{8}}
\put(180,700){\circle*{8}}
\put(280,740){\circle*{8}}
\put( 19,777){\makebox(0,0)[lb]{\smash{-}}}
\put( 17,736){\makebox(0,0)[lb]{\smash{+}}}
\put(266,696){\makebox(0,0)[lb]{\smash{$O(u_R^6)$}}}
\put(117,816){\makebox(0,0)[lb]{\smash{+}}}
\put(360,740){\circle*{8}}
\put(280,780){\circle*{8}}
\put( 60,700){\oval( 40, 20)[tr]}
\put( 60,700){\oval( 40, 20)[tl]}
\put(110,740){\oval( 20, 10)[tr]}
\put(110,740){\oval( 20, 10)[tl]}
\put( 90,740){\oval( 20, 10)[tr]}
\put( 90,740){\oval( 20, 10)[tl]}
\put( 70,740){\oval( 20, 10)[tr]}
\put( 70,740){\oval( 20, 10)[tl]}
\put(100,700){\oval( 40, 20)[tr]}
\put(100,700){\oval( 40, 20)[tl]}
\put(200,700){\oval( 80, 20)[tr]}
\put(200,700){\oval( 80, 20)[tl]}
\put(220,740){\oval( 40, 20)[tr]}
\put(220,740){\oval( 40, 20)[tl]}
\put(190,740){\oval( 20, 10)[tr]}
\put(190,740){\oval( 20, 10)[tl]}
\put(170,740){\oval( 20, 10)[tr]}
\put(170,740){\oval( 20, 10)[tl]}
\put( 50,740){\oval( 20, 10)[tr]}
\put( 50,740){\oval( 20, 10)[tl]}
\put(240,820){\oval( 40, 20)[tr]}
\put(240,820){\oval( 40, 20)[tl]}
\put(170,820){\oval( 20, 10)[tr]}
\put(170,820){\oval( 20, 10)[tl]}
\put(150,820){\oval( 20, 10)[tr]}
\put(150,820){\oval( 20, 10)[tl]}
\put( 90,820){\oval( 20, 10)[tr]}
\put( 90,820){\oval( 20, 10)[tl]}
\put( 50,780){\oval( 20, 10)[tr]}
\put( 50,780){\oval( 20, 10)[tl]}
\put(270,780){\oval( 60, 20)[tr]}
\put(270,780){\oval( 60, 20)[tl]}
\put(180,780){\oval( 40, 20)[tr]}
\put(180,780){\oval( 40, 20)[tl]}
\put(150,780){\oval( 20, 10)[tr]}
\put(150,780){\oval( 20, 10)[tl]}
\put( 90,780){\oval( 20, 10)[tr]}
\put( 90,780){\oval( 20, 10)[tl]}
\put( 70,780){\oval( 20, 10)[tr]}
\put( 70,780){\oval( 20, 10)[tl]}
\put(291,740){\oval( 18, 10)[tr]}
\put(291,739){\oval( 20, 12)[tl]}
\put(140,780){\circle*{8}}
\put(100,780){\circle*{8}}
\put( 80,780){\circle*{8}}
\put( 60,780){\circle*{8}}
\put(160,780){\circle*{8}}
\put(260,780){\circle*{8}}
\put(240,780){\circle*{8}}
\put(200,780){\circle*{8}}
\put(180,780){\circle*{8}}
\put( 40,780){\circle*{8}}
\put(100,820){\circle*{8}}
\put( 80,820){\circle*{8}}
\put( 40,820){\circle*{8}}
\put(333,740){\oval( 58, 20)[tr]}
\put(333,739){\oval( 62, 22)[tl]}
\put(140,820){\circle*{8}}
\put(260,820){\circle*{8}}
\put(240,820){\circle*{8}}
\put(220,820){\circle*{8}}
\put(180,820){\circle*{8}}
\put(160,820){\circle*{8}}
\end{picture}
\end{center}

We will further discuss the structure of this series in a later
section.

\subsection{Resummed perturbation theory}\label{resumpert}

For two directed polymers interacting by a short range interaction
(or equivalently
one directed polymer interacting with a straight defect line by a
short range interaction), an exact implicit equation for the
dependence of the free energy per length from the interaction
constant can be given by resummation of the perturbation series.

In order to get these results, we review here the summation technique
given in~\cite{dupl} and generalize it to arbitrary
boundary conditions. After we got the implicit equation we will
study some useful examples of specific boundary conditions.

\subsubsection{Resummation for arbitrary boundary conditions}

The main idea, which leads to the summability, is that the
coefficients of the perturbation series of the partition function have
a product structure, which leads to a simple geometric series if they
are properly decoupled.

This decoupling is achieved by Laplace transforming the constituents
of equation~(\ref{mpointbygs}). We will call them for simplicity
\[
f(t)\equiv G_t(0,0),\quad
{\cal N}\equiv\int{\rm d}r_1\int{\rm d}r_2 G_{L_{||}}(r_1,r_2),
\]
\[
g(t)\equiv\int{\rm d}rG_t(r,0),
\quad\mbox{and}\quad
h(t)\equiv \int{\rm d}rG_t(0,r).
\]
Their Laplace transforms are denoted by $\hat f$, $\hat g$ and
$\hat h$ respectively.

Performing the Laplace transformation on the coefficients of the
partition function, an implicit equation for the free energy per
length can be extracted. The argumentation is given in
appendix~\ref{appimpeqderiv}. The result is that
\begin{equation}
\lim_{L_{||}\to\infty}{F(g)\over L_{||}}=-z_0.
\end{equation}
where $z_0$ is the solution of
\begin{equation}
1+g\hat f(z)=0
\end{equation}
with the largest (absolutely smallest) real part.

This is an exact implicit equation for the free energy per length
derived from perturbation theory.

Introducing again dimension-less coupling constants we get the
equation
\begin{equation}
1+u_0 L^{-\epsilon}\hat f(L^{-1}(-f_0-u_R))=0,
\end{equation}
where $f_0$ denotes the dimension-less free energy per length of the
free ($u_0=0$) problem. As it is obvious from the above equation, it
can be calculated as the smallest pole of the function
$\hat f(-L^{-1}z)$.

To further improve this equation, we write it as
\begin{equation}\label{u0ofur}
1+u_0 F(u_R)=0
\end{equation}
with $F(z)\equiv L^{-\epsilon}\hat f(L^{-1}(-f_0-z))$. Since this
equation has to be fulfilled for $u_0=0$ and $u_R=0$, $F$ has to
behave like $-1/z$ at zero (which can also be verified for specific
infrared regularizations). This means that $H(z)\equiv F(z)+1/z$ is a
regular function at zero. Expressed by $H$ instead of $F$, the
implicit equation for $u_0(u_R)$ reads
\begin{equation}\label{u0ofurreg}
u_R-u_0+u_0u_RH(u_R)=0
\end{equation}

\subsubsection{Results for specific infrared
regularizations and the duality
relation}\label{specresults}\label{duality}

Now we want to calculate $H$ for specific infrared regularizations,
in order to compare the general equation~(\ref{u0ofurreg}) with
results from the literature.

Therefore we first reconsider the definition of $H(z)$
\begin{eqnarray}\label{hdef}
H(z)&=&L^\epsilon \int_0^\infty G_t(0,0) e^{{f_0+z\over L}t}{\rm d}t
+{1\over z}\\\nonumber
&=&\int_0^\infty\left[ L^{1-\epsilon}G_{sL}(0,0)e^{f_0s}-1
\right]e^{zs}{\rm d}s
\end{eqnarray}
{From} dimensional analysis $G_{sL}(0,0)$ must have the form
\begin{equation}
G_{sL}(0,0)=(sL)^{\epsilon-1}g_{IR}(sL).
\end{equation}
Since $G_{sL}(0,0)$ has to decay exponentially with the ground state
energy of the quantum mechanical problem, i.e. like $e^{-sf_0}$, it is
easy to see that $H(z)$ is well-defined for $z\to0$ in the infrared,
because it follows that
\begin{equation}\label{defhir}
L^{(1-\epsilon)}G_{sL}(0,0)e^{f_0s}-1=s^{\epsilon-1}h_{IR}(s)
\end{equation}
with a for large arguments exponentially decaying function
$h_{IR}(z)$.

Also it is clear that ultraviolet divergences arise for
$\epsilon\to0$. The derivatives of $H$ with respect to $z$ are less
divergent in the ultraviolet regime; the $k$-th derivative will
develop ultraviolet divergences at $\epsilon\to-k$. Thus in the case
$\epsilon^\prime\to0$ which we are interested in, also the first
derivative of $H$ at $z=0$ is ultraviolet divergent.

The easiest case is the harmonic regularization, because the full
propagator of the quantum mechanical harmonic oscillator is
analytically known in all dimensions~\cite{qmbook}. From this we
extract the return probability for our directed polymer problem by
inserting ${\bf r}={\bf r}^\prime=0$ and get
\begin{equation}
L^{1-\epsilon}G_{sL}(0,0)=(2\sinh(\pi s))^{-(1-\epsilon)}
\end{equation}
{From} the decay at large $s$ we conclude $f_0=\pi\frac d2$.

The Laplace transform of this function can be explicitly performed
(\cite{gradsteyn} 3.541) and gives
\begin{equation}\label{harmlaplace}
F(z)={\Gamma(\epsilon)\over2\pi}{\Gamma(-{z\over2\pi})\over
\Gamma(\epsilon-{z\over2\pi})}
\end{equation}
For $d=1$ or $\epsilon=\frac12$ this formula together with the
equation~(\ref{u0ofur}) coincides with the one--dimensional transfer
matrix result in~\cite{hiergdiplom}. We want to stress that we just
calculated the full transition function for the free energy from the
Gaussian to the non--Gaussian fixed point in any dimension in the
case of a harmonic boundary condition.

It is remarkable that for $d=3$ or
$\epsilon=-1/2$ the equation~(\ref{u0ofur}) is equivalent to the
$d=1$ case, if one replaces $g$ up to a numerical factor by $-1/g$.
This shows a duality (which exists for all dimensions which are
symmetric with respect to $d=2$): The transition from the Gaussian
fixed point in a dimension below $d=2$ via an repulsive interaction
to the non-Gaussian fixed point is exactly the same as the
transition from the Gaussian fixed point in the symmetric dimension
above $d=2$ to the non--Gaussian fixed point via an attractive
potential.

An explicit calculation of the Laplace transform of the propagator
with hard wall (Dirichlet) boundary conditions
(see appendix~\ref{appdirilap}) in $d=3$ gives the equation
\begin{equation}
\frac8{u_0}=\sqrt{2(u_R+f_0)}\cot(\sqrt{2(u_R+f_0)})
\end{equation}
which shows the same duality relation with the transfer matrix result
for $d=1$ in~\cite{hiergdiplom}.

\subsection{Comparison of the two perturbative approaches}

We will now see that the resummed and the naive perturbation series
after all manipulations presented in the last two sections are
exactly equivalent.

The resummation method gives us the exact relation~(\ref{u0ofurreg})
between the finite size
coefficient $u_R$ of the free energy per length and the dimension-less
coupling constant $u_0$.
If we assume that $H$ can be expanded in a Taylor series around $z=0$,
a perturbation series can be easily extracted from this equation and
it is the same perturbation series as we already calculated directly,
if we identify
\[
a_n\equiv\underbrace{
\setlength{\unitlength}{0.008in}%
\begin{picture}(168,15)(56,796)
\thicklines
\put(160,800){\circle*{8}}
\put(200,800){\circle*{8}}
\put(180,800){\circle*{8}}
\put(220,800){\circle*{8}}
\put( 80,800){\circle*{8}}
\put(140,800){\oval(160, 20)[tr]}
\put(140,800){\oval(160, 20)[tl]}
\put(120,800){\circle*{8}}
\put(100,800){\circle*{8}}
\put( 60,800){\circle*{8}}
\put(140,800){\circle*{8}}
\end{picture}
}_{\hbox{$n$ intervals}}
\equiv {1\over (n-1)!}
\left.{{\rm d}^{n-1}\over{\rm d}z^{n-1}}\right|_0 H(z).
\]
Using the specific representation~(\ref{hdef}) of $H(z)$ by the
function $h_{IR}(z)$ (\ref{defhir}), which is directly connected to
the return probability of the free problem, the coefficients are
calculated as
\begin{equation}\label{acalc}
a_n={1\over (n-1)!}\int_0^\infty s^{\epsilon-1+n-1}h_{IR}(s)
{\rm d}s
\end{equation}

For the comparison with the transfer matrix results, we have to
calculate the free energy per length at the transition point.
Because we did use a renormalization point condition, which gives the
renormalized coupling constant a physical meaning, in this framework,
the finite size coefficient of the free energy per length is just the
fixed point value of the renormalized coupling constant.

\begin{figure}[ht]
\begin{center}
\epsfig{figure=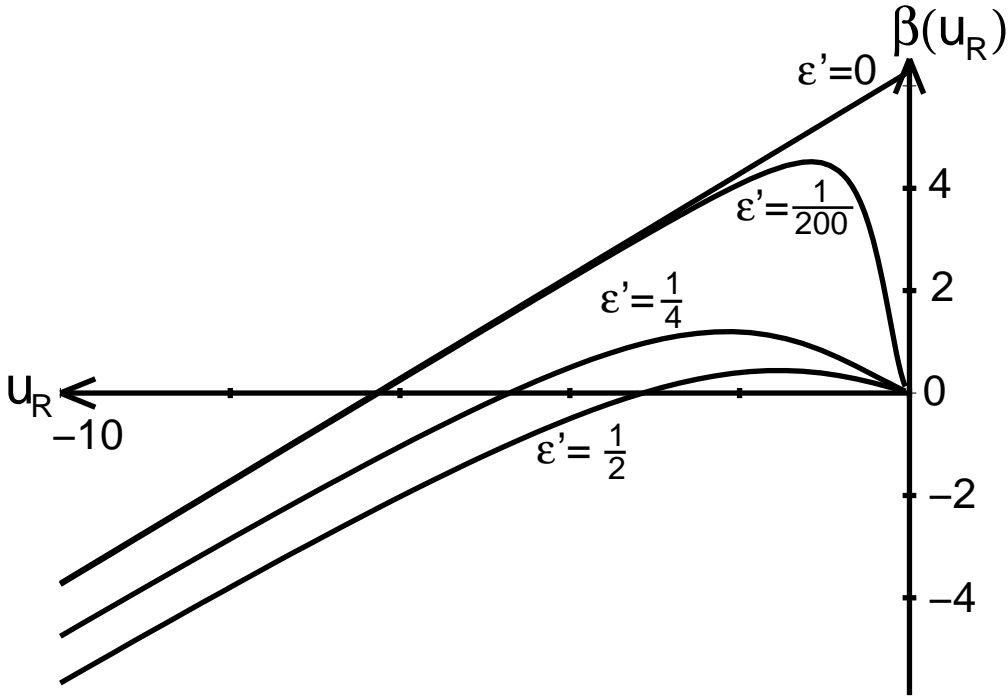}
\caption{The $\beta$ function of a directed polymer in a harmonic
potential.
\label{figharmbeta}}
\end{center}
\end{figure}

This fixed point value is the root of the $\beta$ function, which
describes the flux of the renormalized coupling constant, and which
according to the chain rule is given by
\begin{equation}
\beta(u_R)=\epsilon{u_0(u_R)\over{\partial\over\partial u_R}u_0(u_R)}.
\end{equation}
Using equation~(\ref{u0ofur}) this can be expressed as
\begin{equation}
\beta(u_R)=-\epsilon{F(u_R)\over F^\prime(u_R)}.
\end{equation}
This enables us to calculate this $\beta$-function explicitly in the
case of a harmonic infrared regularization, where the Laplace
transform of the return probability can be
performed~(\ref{harmlaplace}). The result is
\begin{equation}
\beta(u_R)=
{2\pi\epsilon\over\Psi(-{u_R\over2\pi})-\Psi(\epsilon-{u_R\over2\pi})}
\end{equation}
with the digamma function
$\Psi(z)={{\rm d}\over{\rm d}z}\ln \Gamma(z)$.
The behavior of the $\beta$~function already has been shown in
Fig.~\ref{figharmbeta}.
For $\epsilon<0$ there is always one negative root and the slope
of the $\beta$ function at the root is $-\epsilon$. It is remarkable
that the $\beta$ function has a well defined limit function for
$\epsilon\to -1$ ($\epsilon^\prime\to 0$), which is the case we are
mostly interested in. The limiting $\beta$ function is just the
straight line $u_R+2\pi$.

\subsection{The limit $\epsilon^\prime\to 0$}

In the last section, we remarked that despite of the singularities in
the coefficients of the perturbation series, a limiting $\beta$
function for $\epsilon^\prime\to0$ exists. We heavily used the fact
that we could calculate a closed form expression for the
$\beta$ function, and its properties are a bit awkward. Especially
$\beta(0)$ is not zero any more in contrary to all finite orders of
perturbation theory.

If we want to get results for theories, where no exact solution is
possible, we have to find a way, to extract the behavior of the
$\beta$ function in the limit $\epsilon^\prime\to0$ directly from the
lowest order terms of the perturbation expansion.

We will give here first a general description of the regularization
scheme and then show, how it can be applied to reproduce the limiting
$\beta$ function just shown. Moreover we will get results for other
boundary conditions, too.

\subsubsection{General regularization scheme}

Let us assume a perturbation series
\begin{equation}
f_{\epsilon^\prime}(u)=\sum_{n=1}^\infty b_n(\epsilon^\prime)u^n
\end{equation}
the coefficients of which diverge in the limit $\epsilon^\prime\to0$.
Assuming that $\epsilon^\prime=0$ is a pole of $b_n$ of maximal
$n$-th order we use the Laurent expansion of $b_n$
\begin{equation}
b_n(\epsilon^\prime)=
\sum_{k=0}^\infty a_n^{(k)}(\epsilon^\prime)^{-(n-k)}+
c_n(\epsilon^\prime)
\end{equation}
where $\lim_{\epsilon^\prime\to0}c_n(\epsilon^\prime)=0$. We insert
this expansion into the perturbation series and change the order of
summation to
\begin{equation}
f_{\epsilon^\prime}(u)=\sum_{k=0}^\infty\left[
\sum_{n=0}^\infty a_{n+k}^{(k)}
\left(u\over\epsilon^\prime\right)^n\right]u^k+
\sum_{n=1}^\infty c_n(\epsilon)u^n.
\end{equation}
If now the analytic continuations of the functions
\begin{equation}
h_k(x)\equiv\sum_{n=0}^\infty a_{n+k}^{(k)}x^n
\end{equation}
have well defined limits for $x\to\infty$, the limiting form of
$f_{\epsilon^\prime}$ will be
\begin{equation}
f_0(u)=\sum_{k=0}^\infty\left[\lim_{x\to\infty}h_k(x)\right]u^k.
\end{equation}
Moreover using just a finite number of these terms will give us
a systematic expansion of $f_{\epsilon^\prime}$ for small
$\epsilon^\prime$.

To summarize the method, we have to extract just the most divergent
parts of all the coefficients and use them to set up the functions
$h_k$. Those then have to be analytically continued and the limit
$x\to\infty$ to be taken. The limits will give the coefficients of the
limiting functions.

In the following discussion this general approach has to be modified
in some special cases, but the general idea will remain the same.

\subsubsection{The generic case of two directed polymers}

In both perturbative approaches, we have seen that the coefficients
of the perturbation series can all be expressed by some set of
constants $a_1, a_2, \ldots$ which can either be regarded as the
fundamental graphical element of one arc in a diagram or as a
derivative of the function $H(z)$ defined above. The structure of the
perturbation expansion is independent of the infrared regularization
chosen. Different infrared regularizations just result in different
values of the coefficients $a_i$.

In the limit $\epsilon^\prime\to0$ only the first two coefficients
$a_1$ and $a_2$ will diverge, whereas all the higher coefficients
remain regular, as it has been stated in section~\ref{specresults}.
At this point we should remember that in the usual case $\epsilon\to0$
only the coefficient $a_1$ is divergent. So the main difference in
this problem is the existence of two divergent fundamental diagrams
instead of one.

To extract the pole structure, we assume the Laurent expansion of all
the fundamental coefficients:
\begin{equation}
a_k={a_{k,-1}\over\epsilon^\prime}+a_{k,0}+a_{k,1}\epsilon^\prime+
a_{k,2}{\epsilon^\prime}^2+\ldots,
\end{equation}
where $a_{k,-1}=0$ for $k\ge3$.

Since we can calculate the perturbation series up to quite high
orders, we can insert these expansions in the series and study the
behavior of the most divergent terms. It shows up that the correct
series to regularize is the series for $u_0(u_R)\over\epsilon^\prime$,
because this series has a pole of $n$-th order in $\epsilon^\prime$
among its $n$-th order diagrams.

The singular parts happen to have a quite simple structure. All
diagrams of a given order of divergence (order in $u_R$ minus order
of the pole in $\epsilon^\prime$) can be expressed as a geometric
series modified by coefficients polynomial in the running index $n$.
The degree of those polynomials is at most the divergence order
itself, as can be checked to quite high orders. So these series can
be resummed, analytically continued and their limit for $x\to\infty$
can be taken. 

The explicit form of this series can be found in
appendix~\ref{appexpreg}. The regularized series in the limit
$\epsilon^\prime=0$ turns out to be a geometric series, that is
resummed to
\begin{equation}
\lim_{\epsilon^\prime\to0}{u_0(u_R)\over\epsilon^\prime}=
-{1\over a_{1,-1}+a_{2,-1}\,u_R}
\end{equation}
which leads to a linear limiting $\beta$ function of
\begin{equation}
\lim_{\epsilon^\prime\to0}\beta(u_R)={a_{1,-1}\over a_{2,-1}}+u_R.
\end{equation}

The result is quite remarkable, because it states that the
$\beta$ function is linear independent of the infrared regularization
chosen in the limit $\epsilon^\prime\to 0$, which is surely not true
for finite $\epsilon^\prime$.

The result of course reproduces exactly the explicit
form of the $\beta$-function, that we calculated for harmonic boundary
conditions.

It is interesting to state that the coefficients $a_{1,-1}$ and
$a_{2,-1}$ in the limiting $\beta$ function can be calculated much
more easily than all the other coefficients, which were temporally
involved in the calculation. We just have to recall the definition of
$a_1$ and $a_2$ (\ref{acalc}). Integrating by parts the integral for
$a_1$ two times and the integral for $a_2$ one time and dropping the
boundary terms, which are zero at least for $\epsilon>0$, we get
\begin{eqnarray}
a_1&=&{1\over\epsilon^\prime(\epsilon^\prime-1)}\int_0^\infty
s^{\epsilon^\prime}h_{IR}^{\prime\prime}(s){\rm d}s\\
a_2&=&-{1\over\epsilon^\prime}\int_0^\infty s^{\epsilon^\prime}
h_{IR}^\prime(s){\rm d} s.
\end{eqnarray}
Since the divergences in $\epsilon^\prime$ are explicit in these
formulas, we get the coefficients $a_{1,-1}$ and $a_{2,-1}$
by substituting $\epsilon^\prime=0$ in the converging integrals. This
results in
\begin{equation}
a_{1,-1}=h_{IR}^\prime(0)\hbox{ and }a_{2,-1}=h_{IR}(0).
\end{equation}
So these coefficients are directly expressible by the return
probability without any integrations.

\subsubsection{The case of a vanishing leading singularity}

The above derivation has one severe problem. Although the calculated
regularized series (and therefore the $\beta$ function) is well
defined in the limit of $a_{1,-1}\to0$, we relied during the whole
calculation heavily on the fact that $a_{1,-1}$ is not zero.

Unfortunately this is not true in the very important case of
von-Neumann or periodic boundary conditions. These boundary conditions
are the most simple ones and the perturbation series of more than
two directed polymers can only be set up for periodic boundary
conditions.

With periodic boundary conditions the propagator is just the sum
of free propagators connecting equivalent points
\begin{eqnarray*}
G_t({\bf r},{\bf r}^\prime)&=&\left(1\over2\pi t\right)^{d\over2}
\sum_{k\in{\bf Z}^d}
e^{-{({\bf r}-{\bf r}^\prime-kL_\perp)^2\over2t}}\\
&\stackrel{{\bf r}=0,{\bf r}^\prime=0}{=}&
\left({1\over\sqrt{2\pi t}}\sum_{k\in{\bf Z}}
e^{-{k^2 L_\perp^2\over2t}}
\right)^d.
\end{eqnarray*}
Using $L=\sqrt{2\pi}L_\perp^2$ as
it is defined by $\langle\Phi(0)\rangle_0=L^{1-\epsilon}$,
we get (since $f_0=0$)
\begin{equation}
h_{IR}(s)=
\left(\sum_{k\in{\bf Z}}e^{-{k^2\pi\over s}}\right)^d-s^{d\over2}.
\end{equation}
At $d=4$ $h_{IR}(0)=1$, but $h_{IR}^\prime(0)=0$.

Since we know the whole perturbation series, we can just insert
$a_{1,-1}=0$ and repeat our above calculations. The structure of the
poles changes, because now the only divergence resides in the arc,
which spans 2 intervals. This leads to the fact that the order of the
poles in $\epsilon^\prime$ increases only every second order and the
coefficient connected to all poles is $a_{2,-1}$. We take care of
this fact by grouping together all singularities with the same powers
of $a_{2,-1}u_R^2\over\epsilon^\prime$. The structure of the singular
part of this series is then very similar to the previous one and the
regularization scheme can be used in the same way as before. The
explicit form of the important parts of the perturbation series is
shown in appendix~\ref{appexpreg}; in the limit $\epsilon^\prime\to0$
we get the expected result
\begin{equation}\label{u0urwitha1reg}
\lim_{\epsilon^\prime\to0}{u_0(u_R)\over\epsilon^\prime}=
-{1\over a_{2,-1} u_R}+O(u_R^3),
\end{equation}
which is the same as if we had taken the limit $a_{1,-1}\to0$ in our
above result. Therefore also the limiting $\beta$ function is
\begin{equation}
\beta(u_R)=u_R+O(u_R^2)
\end{equation}

\begin{figure}[ht]
\begin{center}
\epsfig{figure=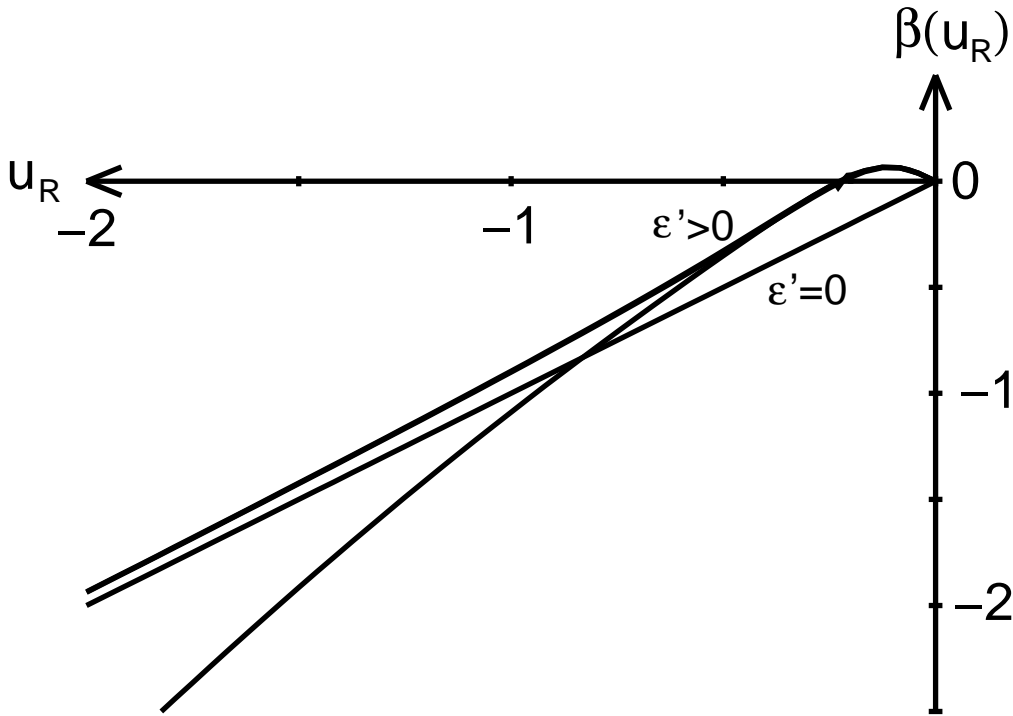}
\caption{Approximations to the \protect$\beta$
function of a directed polymer with
periodic boundary conditions for a small \protect$\epsilon^\prime$.
The straight line is the limiting \protect$\beta$ function
\protect$\beta(u_R)=u_R$ for
\protect$\epsilon^\prime=0$. The other three are calculated with the
same \protect$\epsilon^\prime=0.005$ but truncated at different
orders. Two of them already coincide.\label{figbetafineps}}
\end{center}
\end{figure}

There is one more remarkable thing in this $\beta$ function: If we
just add up the geometric series of the most singular terms and do not
perform the limit
$\epsilon^\prime\to0$, but just insert small $\epsilon^\prime$ we get
the $\beta$ functions shown in Fig.~\ref{figbetafineps}. From this
we can extract two points. First, we can see, how the $\beta$
function with a negative slope at zero approaches its limiting shape
with a slope of $1$ at zero (and everywhere else). Moreover we remark
that all three truncated $\beta$ functions (they differ in the
divergence order where they are truncated) have a common root at
$\pm a_{2,-1}\sqrt{\epsilon^\prime}$. The fact that this is true to
all calculated orders strongly suggests that these roots are exact,
which means that the fixed point approaches zero as
\begin{equation}
u_R\sim\sqrt{\epsilon^\prime},
\end{equation}
which is just equation~(\ref{zeroapp}).

\subsubsection{Comparison between the perturbative and the transfer
matrix approach}

For harmonic boundary conditions, the zero of the $\beta$ function
calculated by the regularized perturbative approaches is exactly
equal to the finite size coefficient calculated in the transfer
matrix picture.

Since the von-Neumann and Dirichlet boundary conditions in the
transfer matrix approach and in the perturbation series are not
absolutely equivalent (in the transfer matrix approach the boundary
has a spherical shape), those results are not quantitatively
comparable. Nevertheless qualitatively also those results are the
same as far as the limit $\epsilon^\prime\to0$ is concerned:

In the case of hard wall (Dirichlet) boundary conditions,
the perturbation series predicts a $\beta$ function with a zero at a
finite negative value. This is also the result of the transfer matrix
calculation due to the finite value of the ground state energy of the
free system.

With periodic (von-Neumann) boundary conditions, the perturbation
theory and the transfer matrix approach both predict that
${\cal C}^*$ approaches
zero as $\epsilon^\prime\to0$ and they both result in the square root
dependence~(\ref{zeroapp}) of $\epsilon^\prime$.

We can therefore conclude that the chosen method of regularization
of the perturbation series reproduces the exact results of the
transfer matrix approach and therefore is a valid regularization
scheme.

\section{Arbitrary number of directed polymers}\label{arbitN}

Now we will develop a diagrammatic expansion of the theory of an
arbitrary number of directed polymers $N$ interacting by short range
interactions. We will see that the structure shows strong analogies
to the case of only two directed polymers,
which suggests that the perturbation series can be regularized in the
same way as the perturbation series for $N=2$.

\subsection{Diagrammatic expansion}

\subsubsection{Perturbation series of the partition function}

In order to keep the calculations as simple as possible, we choose
periodic boundary conditions as the infrared regularization. That
means that we identify the points $r$ and $r+kL_\perp$ for every
$k\in{\bf Z}^d$. Writing down the corresponding propagators, it is
clear that the problem of $N$ directed polymers with periodic
boundary conditions is equivalent to the
problem of $N$ free directed polymers with periodic interactions.

So the perturbation series for the partition function remains the same
as~(\ref{Zseries}) with the definitions
\begin{equation}
\Phi(t)=\left(\sum_{i<j}^N\sum_{k_{ij}\in{\bf Z}^d}
\delta(r_i(t)-r_j(t)-k_{ij}L_\perp)\right)
\end{equation}
and the free expectation values
\widetext
\begin{equation}
\langle\cdot\rangle_0\equiv L_\perp^{-Nd}\!\!\!\!\!
\int\limits_{[0,L_\perp]^d}\!\!\!\!\!{\rm d}^dr_1^\prime\ldots
\!\!\!\!\!
\int\limits_{[0,L_\perp]^d}\!\!\!\!\!{\rm d}^dr_N^\prime
\int\limits_{{\bf R}^d}\!\!{\rm d}^dr_1^{\prime\prime}\ldots\!\!
\int\limits_{{\bf R}^d}\!\!{\rm d}^dr_N^{\prime\prime}
\!\!\!\!
\int\limits_{(r_1^\prime\ldots,r_N^\prime,0)}^{(r_1^{\prime\prime},
\ldots,r_N^{\prime\prime},L_{||})}\!\!\!\!\!\!
{\cal D}r_1\ldots{\cal D}r_N\cdot
e^{-\frac12\int_0^{L_{||}}\sum\limits_{i=1}^N\dot r_i^2(t){\rm d}t}
\end{equation}
\narrowtext
As in the above discussion, we first have to calculate the multi-point
functions in the perturbation series of the partition function and
combine them afterwards to the connected multi-point functions which
are the integrands of the perturbation series of the free energy.

Since the argument of a multi-point function is the sum over different
$\delta$ interactions, all sums can be extracted from the expectation
value. In the $n$-th order the sums over $i<j$ correspond to the
different ways, the $n$ interactions can be arranged among the $N$
directed polymers. Obviously a lot of arrangements are equivalent
since for example interchanging the polymers that interact should not
influence the expectation value of the sequence of
interactions. So every possible arrangement of interactions will be
accompanied by a combinatorial prefactor, which counts the number
of equivalent arrangements. The sum over all $i<j$ can then be written
as a sum over all possible arrangements with each arrangement
multiplied by a convenient combinatorial prefactor.

Since there are obviously $N(N-1)\over2$ possibilities to place the
first interaction, all these prefactors will be multiples of
$N(N-1)\over2$. This is the reason, why we divide the free energy by
this factor in order to get the renormalized coupling constant.

To improve the bookkeeping, we will represent every arrangement of
interactions graphically by drawing parallel lines representing the
directed polymers and a connection between two of them for every
$\delta$ function between two polymers. The first connection belongs
to the ``time'' $t_1$, the last one to the time $t_n$. To write down
some simplified expressions later on, we also introduce a dotted
line, which can represent a ``time'' in the diagram, where no
interaction is present, and which is used to keep the relation
between the time variables and the interactions, if an interaction
is dropped due to a simplification rule.

Between two ``interaction times'' we have to insert the free
propagator for $N$ directed polymers which of course factorizes in a
product of free 1-polymer-propagators
\begin{eqnarray*}
G_{\Delta t}(r^\prime,r^{\prime\prime})&=&
\int_{(r^\prime,0)}^{(r^{\prime\prime},\Delta t)}{\cal D}r
e^{-\frac12\int_0^{\Delta t}\dot r^2(t){\rm d}t}\\
&=&{1\over(2\pi\Delta t)^{\frac d2}}
e^{-{(r^\prime-r^{\prime\prime})^2\over2\Delta t}}
\end{eqnarray*}
Since this is a Gaussian, all integrations over the $r_i$ can be
performed. Moreover it is clear from the translational invariance of
the propagator that only a dependence on ``time'' differences will
arise. Therefore we change our integration variables. Instead of
integrating over all ordered ``times'' $0<t_1<\ldots<t_n<L_{||}$,
we integrate over all existing
``time'' differences $t_i-t_{i-1}$. In the limit of infinitely long
directed polymers
($L_{||}\to\infty$) the domain of integration is from zero to infinity
for each of these variables, whereas the integration over $t_1$ just
gives a factor of $L_{||}$ which is canceled because we want to
calculate the free energy per length and per pair of directed
polymers. The new integration
variables will be made dimension-less by a factor of $L$ and we call
them $s_1,\ldots,s_{n-1}$.

To calculate the integrands, some simplification techniques can be
used and especially two general simplification rules, which we will
call lemma~A and B can be derived. The integrands of the first three
orders of the perturbation series for the partition function can be
explicitly calculated by using these techniques. The detailed
simplification rules and the calculation for the first three orders
can be found in appendix~\ref{app3ordersN} and lead to
\begin{equation}
{2\over N(N-1)}\langle\Phi(t)\rangle_0=L_\perp^{-d},
\end{equation}
\widetext
\begin{equation}\label{secondorderN}
{2\over N(N-1)}\langle\Phi(0)\Phi(Ls_1)\rangle_0=
L_\perp^{-d}R^p(s_1)+{(N-1)(N-2)\over2}L_\perp^{-2d},
\end{equation}
and
\begin{eqnarray}\nonumber
{2\over N(N-1)}\langle\Phi(0)\Phi(Ls_1)\Phi(L(s_1+s_2))\rangle_0
&=&
L_\perp^{-d}R^p(s_1)R^p(s_2)+{(N+1)(N-2)\over2}
L_\perp^{-2d}R^p(s_1)\\\label{thirdorderN}
&&+{(N+1)(N-2)\over2}L_\perp^{-2d}R^p(s_2)\\\nonumber
&&+{(N+5)(N-2)\over2}L_\perp^{-2d}R^p(s_1+s_2)\\\nonumber
&&+{(N-2)(N-3)(N^2+3N+4)\over4}L_\perp^{-3d}
\end{eqnarray}
\narrowtext
with the two-polymer return probability
\begin{equation}
R^p(s)=\left(\frac1{\sqrt{L}}\sum_{k\in{\bf Z}}
e^{-4\pi k^2s}\right)^d.
\end{equation}
We want to stress at this point that up to the third order the only
quantity arising is the two-polymer return probability $R^p(s)$ and
the only effect of more than two directed polymers are the
combinatorial prefactors.
Unfortunately we cannot just stop our discussion here, because in the
forth order, generically new diagrams arise. Since the number of
diagrams heavily increases, it is convenient to use some computer
algebra in order to generate all diagrams. In the sixth order for
example there are 29388 possible arrangements of the interactions,
which can be reduced to 5300 by automatically applying lemma~A.

The new diagrams in the forth order are the following ones
\begin{center}
\setlength{\unitlength}{0.006250in}%
\begin{picture}(380,66)(20,757)
\thicklines
\put(160,780){\line( 1, 0){100}}
\put(300,820){\line( 1, 0){100}}
\put(300,800){\line( 1, 0){100}}
\put(300,780){\line( 1, 0){100}}
\put(160,800){\line( 1, 0){100}}
\put( 20,820){\line( 1, 0){100}}
\put( 20,800){\line( 1, 0){100}}
\put( 20,780){\line( 1, 0){100}}
\put( 20,760){\line( 1, 0){100}}
\put(160,820){\line( 1, 0){100}}
\put( 40,820){\line( 0,-1){ 20}}
\put(340,800){\line( 0,-1){ 20}}
\put(360,820){\line( 0,-1){ 20}}
\put(380,800){\line( 0,-1){ 20}}
\put(100,800){\line( 0,-1){ 40}}
\put( 80,820){\line( 0,-1){ 40}}
\put(320,820){\line( 0,-1){ 20}}
\put( 60,780){\line( 0,-1){ 20}}
\put(180,820){\line( 0,-1){ 20}}
\put(240,820){\line( 0,-1){ 20}}
\put(220,800){\line( 0,-1){ 20}}
\put(200,800){\line( 0,-1){ 20}}
\put(220,780){\circle*{6}}
\put(100,760){\circle*{6}}
\put(100,800){\circle*{6}}
\put(180,820){\circle*{6}}
\put(180,800){\circle*{6}}
\put( 80,820){\circle*{6}}
\put( 40,820){\circle*{6}}
\put( 40,800){\circle*{6}}
\put( 60,760){\circle*{6}}
\put( 60,780){\circle*{6}}
\put( 80,780){\circle*{6}}
\put(200,800){\circle*{6}}
\put(340,780){\circle*{6}}
\put(360,800){\circle*{6}}
\put(360,820){\circle*{6}}
\put(380,800){\circle*{6}}
\put(380,780){\circle*{6}}
\put(340,800){\circle*{6}}
\put(220,800){\circle*{6}}
\put(240,820){\circle*{6}}
\put(200,780){\circle*{6}}
\put(320,820){\circle*{6}}
\put(320,800){\circle*{6}}
\end{picture}
\end{center}
They can of course also be integrated out, which gives
$L_\perp^{-3d}R^p(s_1+2s_2+s_3)$ for the first one. The other two
contain two sums over ${\bf Z}^d$ that cannot be decoupled. The
second one for example reads
\widetext
\begin{equation}
\left\{{1\over4\pi L^2}\left[{\rm det}
\left(\begin{array}{cc}s_1+s_2+s_3&\frac12s_2\\
\frac12s_2&s_2\end{array}\right)
\right]^{-\frac12}\sum_{k,k^\prime\in{\bf Z}}
e^{-{(k\,k^\prime)
\left(\begin{array}{cc}t_1+t_2+t_3&\frac12t_2\\
\frac12t_2&t_2\end{array}\right)
{k\choose{k^\prime}}\over {\textstyle 4\rm det}
\left(\begin{array}{cc}\scriptstyle s_1+s_2+s_3&
\scriptstyle\frac12s_2\\
\scriptstyle\frac12s_2&\scriptstyle s_2\end{array}\right)
}}\right\}^d
\end{equation}
\narrowtext
Here it is already written in a form that makes it obvious that it is
connected to a two-dimensional theta function and that it can be
simplified by applying the corresponding transformation
formula~\cite{bellman} to
\[
\left\{{1\over L^2}\sum_{k,k^\prime\in{\bf Z}}
e^{-4\pi^2{{\textstyle(k\,k^\prime)}
\left(\begin{array}{cc}s_1+s_2+s_3&\frac12s_2\\
\frac12s_2&s_2\end{array}\right)
\left(\begin{array}{c}k\\k^\prime\end{array}\right)
}}\right\}^d
\]
Calculating the diagrams to higher orders shows us that there is a
simple recipe, that allows to perform all spatial integrations
formally, if the form of the diagram is given. The rules are as
follows:
\begin{enumerate}
\item Mark all loops in the diagram that are necessary to pass each
interaction at least with one loop. Assign an orientation to every
loop.
\item If there are $m$ loops necessary, use an $m\times m$ matrix T
and identify each row and column with one of the loops
\item pass for all ``time'' intervals $i$ over all lines of the
diagram and
\begin{itemize}
\item add $\frac12s_i$ to every diagonal element of a loop, that
passes the line
\item add $\pm\frac12s_i$ to both of the off diagonal elements of two
loops, that pass together through the line. If they pass in the same
orientation the plus sign has to be used, otherwise the minus sign.
\end{itemize}
\item The value of the diagram is then
\begin{equation}
\left[L^{-{{\rm order}\over2}}\sum_{k\in{\bf Z}^m}
e^{-4\pi^2 k^TTk}\right]^d.
\end{equation}
\end{enumerate}
This prescription also implies the validity of the lemmas A and B
used above.

Since there are several possibilities to choose the loops in a
diagram, the correspondence from a diagram to a T-matrix is not
unique. For example in the forth order diagram, the loops could be
chosen as:
\begin{center}
\setlength{\unitlength}{0.006250in}%
\begin{picture}(360,86)(20,717)
\thicklines
\put(120,800){\line( 0,-1){ 40}}
\put( 80,760){\line( 0,-1){ 40}}
\put( 40,800){\line( 0,-1){ 40}}
\put(220,720){\line( 1, 0){160}}
\put(220,760){\line( 1, 0){160}}
\put(220,800){\line( 1, 0){160}}
\put( 20,720){\line( 1, 0){160}}
\put(160,760){\line( 0,-1){ 40}}
\put(290,735){\line( 0, 1){ 10}}
\multiput(290,735)(0.41667,-0.41667){13}{\makebox(0.4444,0.6667){.}}
\put(345,730){\vector(-1, 0){ 50}}
\multiput(345,730)(0.41667,0.41667){13}{\makebox(0.4444,0.6667){.}}
\put(350,735){\line( 0, 1){ 10}}
\multiput(350,745)(-0.41667,0.41667){13}{\makebox(0.4444,0.6667){.}}
\put(345,750){\line(-1, 0){ 50}}
\multiput(295,750)(-0.41667,-0.41667){13}{\makebox(0.4444,0.6667){.}}
\multiput(250,795)(-0.41667,-0.41667){13}{\makebox(0.4444,0.6667){.}}
\put(245,790){\line( 0,-1){ 20}}
\multiput(245,770)(0.41667,-0.41667){13}{\makebox(0.4444,0.6667){.}}
\put(250,765){\line( 1, 0){ 30}}
\multiput(280,765)(0.41667,-0.41667){13}{\makebox(0.4444,0.6667){.}}
\put(285,760){\line( 0,-1){ 30}}
\multiput(285,730)(0.41667,-0.41667){13}{\makebox(0.4444,0.6667){.}}
\put(290,725){\line( 1, 0){ 60}}
\multiput(350,725)(0.41667,0.41667){13}{\makebox(0.4444,0.6667){.}}
\put(355,730){\line( 0, 1){ 20}}
\multiput(355,750)(-0.41667,0.41667){13}{\makebox(0.4444,0.6667){.}}
\put(350,755){\line(-1, 0){ 30}}
\multiput(320,755)(-0.41667,0.41667){13}{\makebox(0.4444,0.6667){.}}
\put(315,760){\line( 0, 1){ 30}}
\multiput(315,790)(-0.41667,0.41667){13}{\makebox(0.4444,0.6667){.}}
\put(250,795){\vector(1, 0){ 60}}
\multiput( 85,750)(0.41667,0.41667){13}{\makebox(0.4444,0.6667){.}}
\put( 90,755){\line( 1, 0){ 60}}
\multiput(150,755)(0.41667,-0.41667){13}{\makebox(0.4444,0.6667){.}}
\put(155,750){\line( 0,-1){ 20}}
\multiput(155,730)(-0.41667,-0.41667){13}{\makebox(0.4444,0.6667){.}}
\put( 90,725){\vector( 1, 0){ 60}}
\multiput( 90,725)(-0.41667,0.41667){13}{\makebox(0.4444,0.6667){.}}
\put( 85,730){\line( 0, 1){ 20}}
\multiput( 50,795)(-0.41667,-0.41667){13}{\makebox(0.4444,0.6667){.}}
\put( 45,790){\line( 0,-1){ 20}}
\multiput( 45,770)(0.41667,-0.41667){13}{\makebox(0.4444,0.6667){.}}
\put( 50,765){\line( 1, 0){ 60}}
\multiput(110,765)(0.41667,0.41667){13}{\makebox(0.4444,0.6667){.}}
\put(115,770){\line( 0, 1){ 20}}
\multiput(115,790)(-0.41667,0.41667){13}{\makebox(0.4444,0.6667){.}}
\put( 50,795){\vector(1, 0){ 60}}
\put(240,800){\line( 0,-1){ 40}}
\put(280,760){\line( 0,-1){ 40}}
\put(320,800){\line( 0,-1){ 40}}
\put(360,760){\line( 0,-1){ 40}}
\put( 20,760){\line( 1, 0){160}}
\put(160,720){\circle*{6}}
\put(160,760){\circle*{6}}
\put(120,800){\circle*{6}}
\put(120,760){\circle*{6}}
\put( 40,800){\circle*{6}}
\put( 40,760){\circle*{6}}
\put( 80,760){\circle*{6}}
\put( 80,720){\circle*{6}}
\put(240,760){\circle*{6}}
\put( 20,800){\line( 1, 0){160}}
\put(360,720){\circle*{6}}
\put(360,760){\circle*{6}}
\put(320,800){\circle*{6}}
\put(240,800){\circle*{6}}
\put(280,760){\circle*{6}}
\put(280,720){\circle*{6}}
\put(320,760){\circle*{6}}
\end{picture}
\end{center}
which leads to the T-matrices
\[
\left(\begin{array}{cc}t_1+t_2&\frac12 t_2\\\frac12 t_2&t_2+t_3
\end{array}\right)\quad\hbox{and}\quad
\left(\begin{array}{cc}t_1+t_2+t_3&\frac12 t_2+t_3\\
\frac12 t_2+t_3&t_2+t_3
\end{array}\right).
\]
Although the T-matrices are not identical, the sums, which determine
the value of the T-matrix are the same, because the second of those
matrices is reproduced, if in the sum with the first matrix, the
summation variable $k_2$ is shifted by $k_1$. During all those
equivalence transformations, the determinant of the T-matrix does not
change, so that two T-matrices from different diagrams can only be
equivalent, if their determinants are identical. Unfortunately there
exist also T-matrices which are not equivalent but have the same
determinant.

Since the whole procedure described above is purely mechanical, it
can be implemented by computer algebra. The only thing left to do is
manually checking the equivalence of T-matrices which have the same
determinant.
This reduces the number of different diagrams in the fifth order from
348 to 88.

\subsubsection{Perturbation series of the free energy}

Since we now know how to compute the multi-point functions, we
can calculate the free energy using the formula from
appendix~\ref{appexpansions}. Analogous to the connected two point
functions we used for the 2 polymer problem, we have to express
everything in terms of correlation functions, that decay properly for
increasing time differences. In analogy to the definition of $g_1$,
which in our terminology reads $g_1(s)=L_\perp^dR^p(s)-1$, we will
use functions of T-matrices, where all sub-matrices of lower
dimensions are subtracted with alternating signs, as for example
\begin{eqnarray*}
g_2(a,b,c)&\equiv&\left\{\sum_{k\in{\bf Z}^2} e^{-4\pi^2k^T
\left(\begin{array}{cc}a&b\\b&c\end{array}\right)k}
\right\}^d\\
&&
-\left\{\sum_{k\in{\bf Z}}e^{-4\pi^2ak^2}\right\}^d
-\left\{\sum_{k\in{\bf Z}}e^{-4\pi^2ck^2}\right\}^d+1,
\end{eqnarray*}
and analogously for a higher number of loops.

Of course the subtracted terms depend on the T-matrix itself and are
different, if one chooses different T-matrices which represent the
same diagram (which give the same value for the first term of the
above sum). If one chooses the wrong representation, one will
discover that there are terms, the time integrals of which will not
converge. But it is
always possible to choose a representation, that leads to terms which
are each integrable (of course the sum of all terms is integrable in
every case, it is just possible that one chooses an inconvenient
partition of the integrand.)

Once the coefficients of the free energy are represented as a sum over
time integrals, further simplifications can be performed, by
multiplying the integration variables by constant factors and
applying all kinds of relations as the ones discussed in
appendix~\ref{appdiagrels}.

We will continue to represent every of
those integrals by one of the diagrams, which is responsible for the
main term of the integrand. Since the integrands themselves depend
on the way, the loops are chosen, we must specify our choice.
Those parts of diagrams, which can be expressed only by $g_1$
(i.e. which are pure 2-polymer-diagrams) will be represented
by the same diagrams as they have been used in the 2 polymer case.

\begin{figure}[ht]
\begin{center}
\epsfig{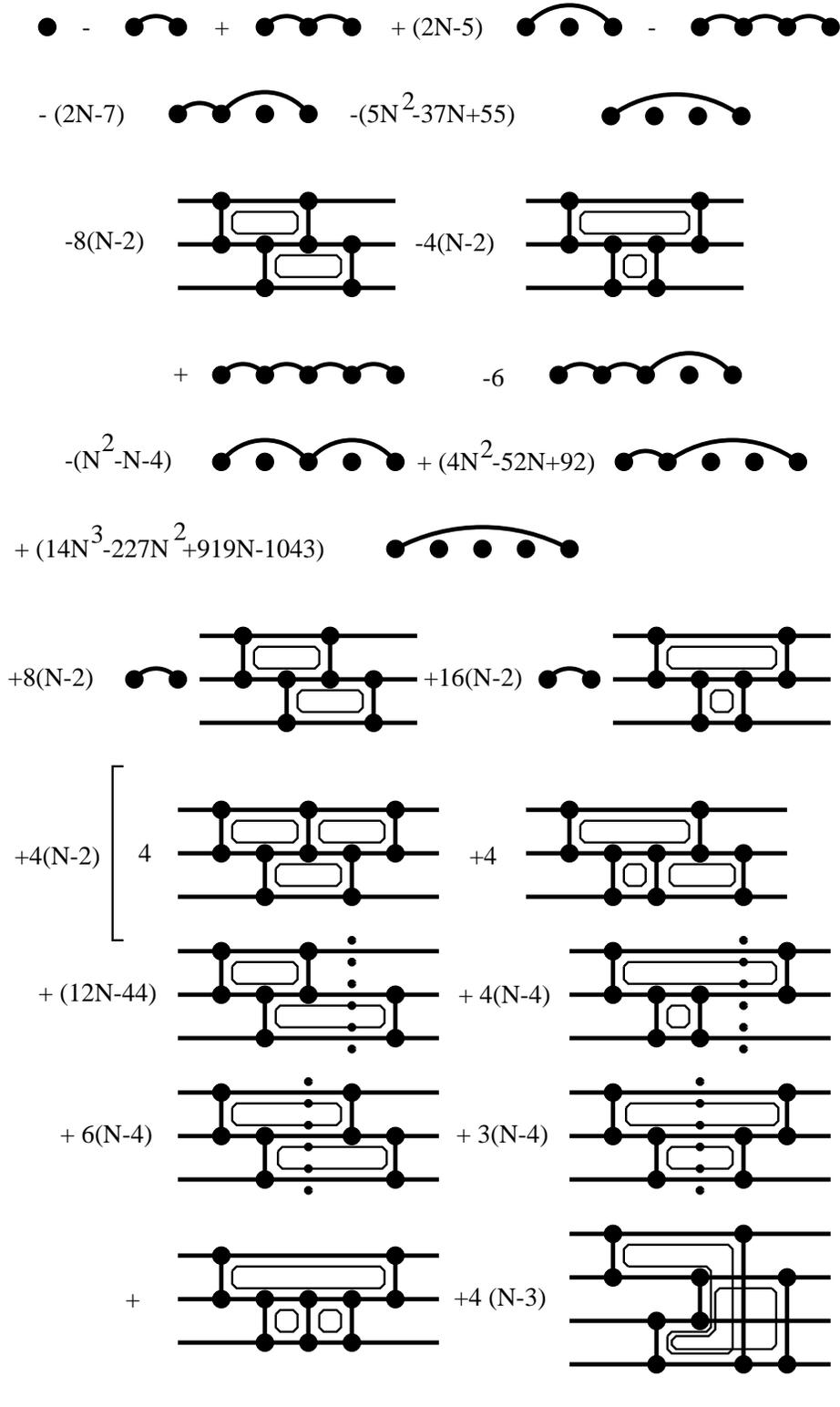}
\caption{Diagrammatic series expansion of the free energy per length
up to the fifth order for an arbitrary number $N$ of interacting
polymers.\label{figNfreener}}
\end{center}
\end{figure}

The resulting diagrammatic expansion for the free energy is shown
in Fig.~\ref{figNfreener} up to the fifth order.

It should be stressed that the number of diagrams involved in this
final expansion is very small compared to the number of diagrams
in the original perturbation series of the partition function.
Since the applicability of most of the simplifying formulas as
those from appendix~\ref{appdiagrels} relies on very specific
relations
between the coefficients of the diagrams involved, it is very unlikely
that this structure evolves just by chance. This is a strong hint that
there exists a similar equation as~(\ref{u0ofurreg}) for an arbitrary
number of directed polymers, that produces this relatively simple
structure of the perturbation series. Unfortunately no such formula
could be found up to now.

\subsection{Comparison with the $N=2$ case}

We can now compare the perturbation series to the perturbation series
of two directed polymers.

First there are a lot of diagrams which appear already in the series
of two directed polymers, that now have prefactors with a polynomial
$N$ dependence.
We will call them ``linear'' diagrams, because they have no nested
loops. The value of one arc itself is of course the same as for two
directed polymers.
Especially only the arc, that spans two intervals has a pole at
$\epsilon^\prime\to0$. This pole is of the first order.

Since we have explicit expressions for all of the diagrams, it is
also possible to extract at least the order of the pole of each
of the nested diagrams as $\epsilon^\prime\to0$. This can be done
by extracting the UV dependence of the integrands and partial
integration. It comes out that all nested diagrams are less divergent
than the product of two-interval arcs in the same order of the
perturbation theory.

Therefore, as in the case of two directed polymers, the powers of the
two-interval arc are the most divergent diagrams and the order of the
pole grows by one only in every second order of the perturbation
series.

Although we gave here the diagrammatic expansion of the series
$u_R(u_0)$, the same statements are true for the inverted series
$u_0(u_R)$, since
it consist of exactly the same diagrams with other prefactors.

Analogous to the case of two directed polymers, we write the partial
sum over all those most divergent diagrams as
\begin{equation}
{u_0(u_R)\over\epsilon^\prime}={1\over u_R}f_{reg}
\left(u_R\over\sqrt{\epsilon^\prime}\right)+
\mbox{less divergent terms}
\end{equation}
where $f_{reg}$ is an even function, with Taylor coefficients, that
stay regular as $\epsilon^\prime\to0$. If we now assume that the fact
that the series for $f_{reg}(x)$ can be analytically continued to
$x\to\infty$ and that $\lim_{x\to\infty}f_{reg}(x)$ exists is not a
particularity of 2 directed polymers, we get a leading behavior of
\begin{equation}
{u_0(u_R)\over\epsilon^\prime}\approx{a\over u_R}\hbox{ with }
a=\lim\limits_{x\to\infty}f_{reg}(x)
\end{equation}
Differentiation then leads to the limiting $\beta$ function
\begin{equation}
\beta(u_R)=u_R+\hbox{higher order terms.}
\end{equation}
The slope of $1$ of this $\beta$ function shows that $u_R=0$ is not
just the Gaussian fixed point, which would have a slope of
$\epsilon=-1$. Thus we expect that for $\epsilon^\prime\to0$ the
non-Gaussian fixed point
goes to zero for each $N$. Since the $\beta$ function is a regular
function of $u_R/\sqrt{\epsilon^\prime}$ also the behavior
proportional to $\sqrt{\epsilon^\prime}$ is independent of $N$.

\subsection{Consequences for the limit $N\to0$}

The independence of the behavior of the fixed point near $d=4$ renders
the limit $N\to0$, which describes the directed polymer in a random
medium, trivial.
Thus we easily conclude that the finite size amplitude of the free
energy per unit ``time'' also vanishes as given in
equation~(\ref{zeroapp}) for the directed polymer in a random medium.

This shows that the singularities in the perturbation series, which
arise at $d=4$ are not only a formal problem of the approach, but
they lead to a non-analytic behavior of a physical quantity.

Although we are surely not able to access the strong coupling fixed
point of the KPZ equation with these perturbative methods, the
shown vanishing of the weak coupling fixed point, which is under the
Hopf--Cole transformation equivalent
to the studied interaction fixed point in the directed polymer
picture, is a strong
hint that $4$ is really the upper critical dimension for the
KPZ problem.

Although the duality relations between dimensions $d$ and $4-d$ in
section~\ref{duality} only have been observed in the case $N=2$,
one could speculate that such a relationship still exists for
arbitrary $N$ and the KPZ problem. This is especially interesting
since Frey and T{\"a}uber find in~\cite{frey94,frey95} that their
coupling constant approaches zero in the limit $d\to0$. Surely it is
not absolutely clear how the two coupling constants are related and
this point deserves further investigations.

\section{Conclusion and outlook}\label{conclusion}

We have studied the problem of directed polymers with short range
interactions. The main interest was the behavior of the directed
polymer system
in the limit $d\to4$, where the normal renormalization treatment,
which expands around $d=2$ breaks down. Since for two directed
polymers exact transfer
matrix results are available, we developed a regularization procedure
for the perturbation series in the limit $d\to4$ in the special case
of two directed polymers. We could prove that there is an agreement
between the transfer matrix results and the predictions of the
regularized perturbation series for two directed polymers.
Moreover an exact implicit
equation which generates the whole perturbation series and its
connections to the diagrammatic expansion has been studied.

For an arbitrary number of directed polymers, a diagrammatic
expansion of the free energy up to the fifth order has been
established. Although it stems from a large number of terms this
expansion is relatively simple. This suggests that it can also be
generated by a simple equation, which has, however, eluded us so far.
Since the pole structure of
the perturbation series is very similar to the pole structure in the
case $N=2$, the leading behavior of the $\beta$ function in the limit
$d\to4$ has been derived. It shows that the interaction fixed point
approaches zero as $d\to4$ proportional to $\sqrt{\epsilon^\prime}$
independent of $N$. Therefore the
weak coupling fixed point of the KPZ equation (which is mapped to the
problem of a directed polymer in a random medium) is also expected
to approach 0 proportional to $\sqrt{\epsilon^\prime}$ in the limit
$d\to4$.

For $d>4$ the transfer matrix calculations show for two directed
polymers, that the finite size amplitude of the free energy per unit
``time'' at the unbinding transition develops anomalous scaling
behavior analogous to that of the free energy in the bound state
of the infinite system~\cite{lipo91,lipo91a}.
The finite size amplitude therefore depends explicitly on the
short-distance cutoff. Thus
it is no longer possible to define a universal quantity from the free
energy. If this behavior persists for arbitrary $N$,
the critical behavior of the KPZ equation at the roughening transition
above $d=4$ is less universal than below 4 dimensions.

We conclude by noting that the unbinding transition of
semi-flexible polymers (i.e. directed lines which are governed
by the curvature
instead of a line tension) with a short-ranged attraction of parallel
polymer segments in one transversal dimension formally can be mapped
on the problem of two strings in four transversal dimensions.
Therefore the method studied here can also be applied to examine this
unbinding transition~\cite{ourpaper}.

\section*{Acknowledgments}

We gratefully acknowledge useful discussions with E. Frey,
C. Hiergeist, R. Lipowsky and U. T{\"a}uber.

\appendix

\section{Solutions of the radial Schroedinger equation with constant
potential}\label{appschroedsol}

We have to calculate the asymptotic behavior of the ground state
energy, which belongs to the rescaled version
\[
\left\{-{\partial^2\over\partial y^2}+{(d-3)(d-1)\over 4}{1\over y^2}+
\bar V(y)\right\}\phi(y)=E_0\phi(y)
\]
of the radial equation~(\ref{radsgl}). The rescaled potential is
\begin{equation}
\bar V(r)=
\left\{\begin{array}{rl}-V_0&0\le r<1\\0&1\le r\end{array}\right.,
\end{equation}
where $\phi(r)=\Phi(ra)$, $E_0=a^2\tilde E_0$, $\bar V(r)=a^2V(ra)$
and $V_0=a^2\tilde V_0$.

Obviously the wave function must consist of general solutions of the
Schroedinger equation with constant potential.
\begin{equation}
\left(-{\partial^2\over\partial y^2}+{(d-3)(d-1)\over 4}{1\over y^2}
-E\right)
\phi(y)=0
\end{equation}
The general solution of this differential equation is~\cite{gradsteyn}
\[
\phi(y)=\left\{\begin{array}{cc}
A\sqrt y J_{|\epsilon|}(y\sqrt E)+B\sqrt y N_{|\epsilon|}(y\sqrt E)&
E>0\\
Ay^{d-1\over2}+By^{3-d\over2}&E=0\\
A\sqrt y I_{|\epsilon|}(y\sqrt{-E})+B\sqrt yK_{|\epsilon|}(y\sqrt{-E})
&E<0
\end{array}\right..
\]

Because the wave function should be regular at the origin at least
for $d>2$ only the first solution is possible for the interaction
region. This means
that for $y\le1$
\begin{equation}
\phi(y)=\left\{\begin{array}{cc}
A\sqrt yJ_{|\epsilon|}(y\sqrt{E_0+V_0})&E_0+V_0>0\\
A\sqrt y y^{|\epsilon|}&E_0+V_0=0\\
A\sqrt y I_{|\epsilon|}(y\sqrt{-E_0-V_0})&E_0+V_0<0\end{array}\right..
\end{equation}
In the outer region the boundary conditions at $y=L/a$ lead to the
wave functions
\[
\phi(y)=
B[\sqrt y J_{|\epsilon|}(y\sqrt{E_0})-
g_{|\epsilon|}(\sqrt{E_0}\frac La)\sqrt yN_{|\epsilon|}(y\sqrt{E_0})]
\]
for $E_0>0$,
\[
\phi(y)=\left\{\begin{array}{ll}
B[y^{d-1\over2}-\left(L\over a\right)^{d-2}y^{3-d\over2}]&
\mbox{ Dirichlet}\\
By^{d-1\over2}&\mbox{ von-Neumann}\end{array}\right.
\]
for $E_0=0$, and
\[
\phi(y)\!=\!B[\sqrt y I_{|\epsilon|}(y\sqrt{-E_0})-
f_{|\epsilon|}(\sqrt{-E_0})\sqrt y K_{|\epsilon|}(y\sqrt{-E_0})]
\]
for $E_0<0$ respectively with
\begin{equation}
f_{|\epsilon|}(x)=
\left\{\begin{array}{ll}I_{|\epsilon|}(x)\over
K_{|\epsilon|}(x)&\hbox{Dirichlet}\\
-{I_{|\epsilon|+1}(x)\over K_{|\epsilon|+1}(x)}&
\hbox{von-Neumann}\end{array}\right.
\end{equation}
and
\begin{equation}
g_{|\epsilon|}(x)=\left\{\begin{array}{ll}J_{|\epsilon|}(x)\over
N_{|\epsilon|}(x)&\hbox{Dirichlet}\\
{J_{|\epsilon|+1}(x)\over N_{|\epsilon|+1}(x)}&
\hbox{von-Neumann}\end{array}\right..
\end{equation}

The total solution consists of these two solutions if they obey
the condition of continuous differentiability, which is most
conveniently written as
\[
\phi(1-)=\phi(1+)
\]
and
\[
\phi^\prime(1-)-(|\epsilon|+\frac12)\phi(1-)=
\phi^\prime(1+)-(|\epsilon|+\frac12)\phi(1+).
\]
Since this are two conditions for the two free coefficients $A$ and
$B$, we only get a solution for energies, where the determinant of
the linear system of equation vanishes. This gives us the equation
connecting $E_0$ with $V_0$ and $L_\perp\over a$.

This equation will always have the form
\begin{equation}
r(\sqrt{\pm E_0}\frac{L_\perp}a)=s(\sqrt{\pm E_0})
\end{equation}
with some functions $r$ and $s$. Since the ground state energy is
expected to vanish for $L_\perp\to\infty$ at least proportional to
$(a/L_\perp)^2$,
we have to distinguish two cases which are determined by the value of
$s_0\equiv\lim_{x\to0}s(x)$. If $s_0$ is a finite value, $E_0$
will asymptotically decay as $\pm c^2(a/L_\perp)^2$, where
$c$ is the smallest positive solution of $r(x)=s_0$. So $c^2$ is
exactly the coefficient, we are interested in. If $s_0=0$, the
constant $c$ is a root of $r(x)$. Since $r(0)=0$, $c$ can be zero,
but only if the signs of $r$ and $s$ for small positive arguments are
equal. If not, $c$ is the smallest positive root of $r$. If $c$ is
zero, one can expand $r$ for small arguments and furthermore extract
the leading behavior in $a/L_\perp$ of $E_0$, which then decays
faster than quadratic.

Since we are especially interested in the asymptotic behavior of the
ground state energy at the phase transition point $V_*$, we have to
identify it. It is defined by the condition that the ground state
energy in an infinite system approaches zero from below. Since we
know the solutions of the Schroedinger equation in the interaction
region and outside (for $L_\perp\to\infty$ only the regular solution
proportional to $K_{|\epsilon|}$
is possible), we can exploit the matching conditions for $E_0+V_*>0$
and $E_0<0$ and take the limit $E_0\to0$, which gives
\begin{equation}
{J_{|\epsilon|+1}(\sqrt{V_*})\sqrt{V_*}\over
J_{|\epsilon|}(\sqrt{V_*})}=2|\epsilon|.
\end{equation}

We now can systematically apply the above scheme to all possible
combinations $V_0=0$, $0<V_0<V_*$ and $V_0=V_*$, $E_0<-V_0$,
$E_0=V_0$, $-V_0<E_0<0$, $E_0=0$ and $E_0>0$ and both boundary
conditions and extract all possible values of the coefficient $c$.
The smallest one is the ground state energy.

The only point, which has to be handled with care during this
calculation is the series expansion of the different Bessel functions
involved. Since especially for $V_0=V_*$ the leading terms of the
expansions cancel, the sub-leading terms have to be used. But the
sub-leading terms are of totally different origin if $|\epsilon|>1$
or $|\epsilon|<1$. This produces the difference of the ground state
energies in $d>4$ and $d<4$ as we expect them.

\section{Expansion of the free energy}\label{appexpansions}

Since we need it during the calculations, we give here the expansion
of the free energy per length written by $n$ point functions up to
the forth order. It is:
\widetext
\begin{eqnarray*}
\lim\limits_{L_{||}\to\infty}\!\!{F(g)-F(0)\over L_{||}}&=&
g\langle\Phi(0)\rangle_0-g^2\int_0^\infty
\langle\Phi(0)\Phi(t)\rangle_0
-\langle\Phi(0)\rangle_0^2{\rm d}t+\\
&+&g^3\int\int\limits_{\llap{$\scriptstyle0\le t_1\le t_2$ }}
{\rm d}t_1
{\rm d}t_2\Bigg[\langle\Phi(0)\Phi(t_1)\Phi(t_2)\rangle_0-
\langle\Phi(0)\rangle_0\langle\Phi(t_1)\Phi(t_2)\rangle_0\\
&&\qquad-\langle\Phi(t_1)\rangle_0\langle\Phi(0)\Phi(t_2)\rangle0-
\langle\Phi(t_2)\rangle_0\langle\Phi(0)\Phi(t_1)\rangle_0
+2\langle\Phi(0)\rangle_0^3\Bigg]\\
&-&g^4\!\!
\int\int\int\limits_{\llap{$\scriptstyle0\le t_1\le t_2\le t_3$}}
{\rm d}t_1{\rm d}t_2{\rm d}t_3\Bigg[
\langle\Phi(0)\Phi(t_1)\Phi(t_2)\Phi(t_3)\rangle_0-\!
\langle\Phi(0)\Phi(t_1)\rangle_0\langle\Phi(t_2)\Phi(t_3)\rangle_0\\
&&\qquad\qquad-\langle\Phi(0)\Phi(t_2)\rangle_0
\langle\Phi(t_1)\Phi(t_3)\rangle_0
-\langle\Phi(0)\Phi(t_3)\rangle_0\langle\Phi(t_1)\Phi(t_2)\rangle_0\\
&&\qquad\qquad-\langle\Phi(0)\rangle_0
\langle\Phi(t_1)\Phi(t_2)\Phi(t_3)\rangle_0
-\langle\Phi(0)\rangle_0
\langle\Phi(0)\Phi(t_2)\Phi(t_3)\rangle_0\\
&&\qquad\qquad
-\langle\Phi(0)\rangle_0\langle\Phi(0)\Phi(t_1)\Phi(t_3)\rangle_0
-\langle\Phi(0)\rangle_0
\langle\Phi(0)\Phi(t_1)\Phi(t_2)\rangle_0\\
&&\qquad\qquad+2\langle\Phi(0)\Phi(t_1)\rangle_0
\langle\Phi(0)\rangle_0^2
+2\langle\Phi(0)\Phi(t_2)\rangle_0\langle\Phi(0)\rangle_0^2\\
&&\qquad\qquad
+2\langle\Phi(0)\Phi(t_3)\rangle_0\langle\Phi(0)\rangle_0^2
+2\langle\Phi(t_1)\Phi(t_2)\rangle_0
\langle\Phi(0)\rangle_0^2\\
&&\qquad\qquad
+2\langle\Phi(t_1)\Phi(t_3)\rangle_0\langle\Phi(0)\rangle_0^2
+2\langle\Phi(t_2)\Phi(t_3)\rangle_0\langle\Phi(0)\rangle_0^2
-6\langle\Phi(0)\rangle_0^4\Bigg]\\
&+&O(g^5).
\end{eqnarray*}
\narrowtext

In the case of only two directed polymers, this whole series can be
reexpressed only by connected two point functions and then represented
diagrammatically as explained in section~\ref{sectdiag2}. It reads
then:
\begin{center}
\setlength{\unitlength}{0.008in}%
\begin{picture}(424,102)(-30,709)
\thicklines
\put(100,720){\circle*{8}}
\put(120,720){\circle*{8}}
\put(160,720){\circle*{8}}
\put(180,720){\circle*{8}}
\put(200,720){\circle*{8}}
\put(220,720){\circle*{8}}
\put( 80,720){\circle*{8}}
\put(200,760){\circle*{8}}
\put(220,760){\circle*{8}}
\put(260,760){\circle*{8}}
\put(280,760){\circle*{8}}
\put(300,760){\circle*{8}}
\put(320,760){\circle*{8}}
\put( 60,720){\circle*{8}}
\put(260,720){\circle*{8}}
\put(-30,796){\makebox(0,0)[lb]{\smash{$u_R(u_0) =$}}}
\put(235,755){\makebox(0,0)[lb]{\smash{+}}}
\put( 40,715){\makebox(0,0)[lb]{\smash{+}}}
\put(135,715){\makebox(0,0)[lb]{\smash{+}}}
\put(238,715){\makebox(0,0)[lb]{\smash{-}}}
\put(332,715){\makebox(0,0)[lb]{\smash{$+O(g^5)$}}}
\put(135,755){\makebox(0,0)[lb]{\smash{+}}}
\put(280,720){\circle*{8}}
\put(300,720){\circle*{8}}
\put(320,720){\circle*{8}}
\put( 62,795){\makebox(0,0)[lb]{\smash{-}}}
\put(115,795){\makebox(0,0)[lb]{\smash{+}}}
\put(198,795){\makebox(0,0)[lb]{\smash{-}}}
\put( 45,755){\makebox(0,0)[lb]{\smash{-}}}
\put(180,760){\circle*{8}}
\put(200,760){\oval( 40, 20)[tr]}
\put(200,760){\oval( 40, 20)[tl]}
\put(280,760){\oval( 40, 20)[tr]}
\put(280,760){\oval( 40, 20)[tl]}
\put(310,760){\oval( 20, 10)[tr]}
\put(310,760){\oval( 20, 10)[tl]}
\put( 90,720){\oval( 60, 20)[tr]}
\put( 90,720){\oval( 60, 20)[tl]}
\put( 90,720){\oval( 20, 10)[bl]}
\put( 90,720){\oval( 20, 10)[br]}
\put(180,720){\oval( 40, 20)[tr]}
\put(180,720){\oval( 40, 20)[tl]}
\put(170,760){\oval( 20, 10)[tr]}
\put(170,760){\oval( 20, 10)[tl]}
\put(240,800){\oval( 40, 20)[tr]}
\put(240,800){\oval( 40, 20)[tl]}
\put( 90,800){\oval( 20, 10)[tr]}
\put( 90,800){\oval( 20, 10)[tl]}
\put(150,800){\oval( 20, 10)[tr]}
\put(150,800){\oval( 20, 10)[tl]}
\put(170,800){\oval( 20, 10)[tr]}
\put(170,800){\oval( 20, 10)[tl]}
\put( 70,760){\oval( 20, 10)[tr]}
\put( 70,760){\oval( 20, 10)[tl]}
\put( 90,760){\oval( 20, 10)[tr]}
\put( 90,760){\oval( 20, 10)[tl]}
\put(110,760){\oval( 20, 10)[tr]}
\put(110,760){\oval( 20, 10)[tl]}
\put(200,720){\oval( 40, 20)[bl]}
\put(200,720){\oval( 40, 20)[br]}
\put(260,800){\circle*{8}}
\put(240,800){\circle*{8}}
\put( 60,760){\circle*{8}}
\put( 80,760){\circle*{8}}
\put(100,760){\circle*{8}}
\put(120,760){\circle*{8}}
\put(160,760){\circle*{8}}
\put(220,800){\circle*{8}}
\put(290,720){\oval( 60, 20)[tr]}
\put(290,720){\oval( 60, 20)[tl]}
\put( 50,800){\circle*{8}}
\put( 80,800){\circle*{8}}
\put(100,800){\circle*{8}}
\put(140,800){\circle*{8}}
\put(160,800){\circle*{8}}
\put(180,800){\circle*{8}}
\end{picture}
\end{center}

\section{Proving relations among diagrams}\label{appdiagrels}

Every diagram is related to a multiple ``time'' integral over a
product of connected two point functions, the arguments of which are
sums of the different integration variables. If we assume that the
two point function can be Laplace transformed, general rules among
the diagrams can be proved. We will show with one example how this
proceeds. We introduce the Laplace transform $\hat g_1$ of $g_1$.
Then:
\widetext
\[
\int_0^\infty ds_1\int_0^\infty ds_2
\int_0^\infty ds_3g_1(s_1+s_2+s_3)g_1(s_2)+
\int_0^\infty ds_1\int_0^\infty ds_2\int_0^\infty ds_3
g_1(s_1+s_2)g_1(s_2+s_3)=
\]
\begin{eqnarray*}
{1\over(2\pi i)^2}\int_c dz_1\int_c dz_2
\hat g_1(z_1)\hat g_1(z_2) \int_0^\infty ds_1\!
\int_0^\infty ds_2\!
\int_0^\infty ds_3 \Big[\frac12e^{z_1(s_1+s_2+s_3)+z_2s_2}&+&
\frac12e^{z_1s_2+z_2(s_1+s_2+s_3)}\\
&+&e^{z_1(s_1+s_2)+z_2(s_2+s_3)}\Big]
\end{eqnarray*}
\narrowtext

{From} its definition $c$ is a path parallel to the imaginary axis
with positive real part. In order to be able to interchange the
integrations, we have used that the inner integrals exist, which is
of course only true, if the real parts of the $z_i$ are negative. But
since the connected correlation functions decay exponentially for
large arguments, the Laplace transform $\hat g_1$ is still analytic
in some region to the left of the imaginary axis and the integration
contour can be shifted there without changing the value of the
integral.

After doing so, we can evaluate the integrals over the exponentials
and get
\widetext
\begin{eqnarray*}
\ldots\!\!&=&{1\over(2\pi i)^2}\int_c dz_1\int_c dz_2\hat g_1(z_1)
\hat g_1(z2)
\left[-{1\over2z_1^2(z_1+z_2)}-{1\over2z_2^2(z_1+z_2)}
-{1\over z_1z_2(z_1+z_2)}
\right]\\
&=&{1\over(2\pi i)^2}\int_c dz_1\int_c dz_2\hat g_1(z_1)
\hat g_1(z_2)
\left(-{z_1+z_2\over2z_1^2z_2^2}\right)
={1\over(2\pi i)^2}\int_c dz_1\int_c dz_2\hat g_1(z_1)
\hat g_1(z_2)
\left(-{1\over z_1^2z_2}\right)\\
&=&{1\over(2\pi i)^2}\int_c dz_1\int_c dz_2\hat g_1(z_1)
\hat g_1(z_2)
\int_0^\infty ds_1\int_0^\infty ds_2\int_0^\infty ds_3
e^{z_1(s_2+s_3)+z_2 s_1}\\
&=&\int_0^\infty ds_1\int_0^\infty ds_2\int_0^\infty ds_3
g_1(s_1)g_1(s_2+s_3).
\end{eqnarray*}
\narrowtext
The diagrammatic expression for this equation is the one in
section~\ref{sectdiag2}.

\section{Derivation of the exact implicit equation for the free
energy}\label{appimpeqderiv}

Using the abbreviations from section~\ref{resumpert}
the $n$-th order term of the partition function series is expressed
by
\begin{eqnarray*}
\hbox to 20mm{\rlap{$\displaystyle
{1\over{\cal N}}\int_0^{L_{||}}{\rm d}t_n
\int_0^{t_n}{\rm d}t_{n-1}\ldots\int_0^{t_2}{\rm d}t_1$}\hfill}&&\\
&&g(t_1)f(t_2-t_1)\ldots f(t_n-t_{n-1})h(L_{||}-t_n)
\end{eqnarray*}
Laplace transforming this with respect to $L_{||}$ yields
$\hat g(z) \hat f(z)^{n-1}\hat h(z)$
for the Laplace transforms.

Obviously the Laplace transformed perturbation series is just a
geometric series and can therefore be resummed. After
back-transformation we end up with
\begin{equation}
{Z\over Z_0}-1=-{g\over2\pi i{\cal N}}\int_c{\hat g(z)\hat h(z)\over
1+g\hat f(z)}e^{L_{||}z}{\rm d}z,
\end{equation}
where $c$ is a path in the complex plane parallel to the imaginary
axis. Since we can obviously close this path by a circle at
$z\to-\infty$, the integral is given as the sum of the residues of
the integrand in the half plane of negative real parts. From the form
of the integrand it is clear
that all residues will be some prefactor times an exponential with
the position of the pole times $L_{||}$ as its argument.
In the limit of $L_{||}\to\infty$ only the pole with the smallest
decay rate (i.e. the one with the smallest absolute value of its
real part) survives.

By construction, it is clear that ${\cal N}$ has a leading dependence
on $L_{||}$ of $e^{-L_{||}\ln Z_0}$, whereas the poles of $\hat g$
and $\hat h$ are exactly the negative eigenvalues of the Schroedinger
operator corresponding to the free directed polymer problem
(described by ${\cal H}_0$).
The smallest eigenvalue is the leading term of $\ln Z_0$ itself,
the contribution of which to
the integral must cancel against the $1$ on the left hand side.

{From} that we conclude that the leading term of $Z$ for large
$L_{||}$ is some prefactor times $\exp(z_0L_{||})$, where $z_0$ is the
solution of the equation
\begin{equation}
1+g\hat f(z)=0
\end{equation}
with the largest (absolutely smallest) real part. This decay rate is
therefore the leading contribution to the free energy per length in
the limit $L_{||}\to\infty$.

\section{Hard wall return probability}\label{appdirilap}

To calculate the return probability of a $1+d$-di\-men\-sio\-nal
directed polymer in a round box, we can use the ``quantum mechanical''
expression of the propagator by the eigenfunctions of the
``particle in a box'' problem. The eigenfunctions of the particle in
a box are Bessel functions of the first kind and for $d>2$ we get
with the correct normalization conditions
\widetext
\begin{equation}
G_t({\bf r},{\bf r}^\prime)={2\over L_\perp^2}\sum_{l=0}^\infty
\sum_{J_{{d-2\over2}+l}(\alpha L_\perp)=0}\!\!\!\!\!\!\!\!\!\!
e^{-{\alpha^2\over2}t}
\sum_m\omega^*_{l,m}(\Omega_{\bf r})\omega_{l,m}(\Omega_{\bf r^\prime})
{r^{2-d\over2}J_{{d-2\over2}+l}(\alpha r){r^\prime}^{2-d\over2}
J_{{d-2\over2}+l}(\alpha r^\prime)\over
\left[J_{{d-2\over2}+l+1}(\alpha L_\perp)\right]^2},
\end{equation}
\narrowtext
which for $d=3$ is the heat equation kernel in~\cite{carsjaeg}.

In the limit $r\to0$, which we need for the return probability, only
the $l=0$ terms stay finite. The sum over the for $l=0$ radially
symmetric eigenfunctions of the angular momentum operator is just one
over the surface of the $d$-dimensional unit sphere. Thus the return
probability is
\begin{equation}
{\Gamma(\frac d2)\over\pi^{d/2}L_\perp^2}
\sum_{J_{d-2\over2}(\alpha L_\perp)=0}
{e^{-{\alpha^2\over2}t}\left(\alpha\over2\right)^{d-2}\over
\left[J_{\frac d2}(\alpha L_\perp)\right]^2}.
\end{equation}

Its Laplace transform can formally be calculated term by term, but
since we know that it is ultraviolet divergent for $d>2$ we introduce
a lower cutoff $a$ of the integration, which give
\begin{equation}
{\Gamma(\frac d2)\over\pi^{d/2}(2L_\perp)^{d-2}}e^{-az}
\sum_{J_{d-2\over2}(\alpha)=0}
{\alpha^{d-2}\over\left[J_{\frac d2}(\alpha)\right]^2}
{e^{-{a\over L_\perp^2}{\alpha^2\over2}}\over zL_\perp^2+
{\alpha^2\over2}}.
\end{equation}
In the prefactor the limit $a\to0$ is possible without any
difficulties.

If we now specialize to the case $d=3$, where the roots of the Bessel
function are just the integer multiples of $\pi$, we can insert the
especially simple expressions for the Bessel functions and end up with
\begin{eqnarray*}
\pi\sum_{n=1}^\infty
{e^{-a{\pi^2n^2\over2}}\over1+{2zL_\perp^2\over\pi^2n^2}}&=&
\pi\sum_{n=1}^\infty e^{-a{\pi^2n^2\over2}}\sum_{k=0}^\infty
\left(-{2zL_\perp^2\over\pi^2n^2}\right)^k\\
&=&
\pi\sum_{k=0}^\infty\left(-{2\over\pi^2}\right)^k(zL_\perp^2)^k
\sum_{n=1}^\infty
{e^{-a{\pi^2n^2\over2}}\over n^{2k}},
\end{eqnarray*}
where we have absorbed a factor of $1/L_\perp^2$ into $a$ and omitted
the geometrical prefactor. For $k\ge1$ the limit $a\to0$ is possible
and we get
\begin{eqnarray*}
\hbox to 5mm{\rlap{$\displaystyle
\pi\sum_{n=1}^\infty e^{-a{\pi^2n^2\over2}}+
\pi\sum_{k=1}^\infty\left(-{2\over\pi^2}\right)^k
\zeta(2k)(zL_\perp^2)^k
$}\hfill}&&\\
&=&\pi\sum_{n=1}^\infty e^{-a{\pi^2n^2\over2}}-{\pi\over2}
\left[L_\perp\sqrt{-2z}\cot(L_\perp\sqrt{-2z})-1\right].
\end{eqnarray*}
The last equation can be verified by representing the values of the
$\zeta$ function by Bernoulli numbers which gives exactly the series
expansion of the cotangent function. The $k=0$ sum obviously diverges
for $a\to0$. If we add $\frac\pi2$, it is half of the value of theta
function at zero, which diverges like $a^{-\frac12}$ with no
sub-leading algebraic terms. So, if we ignore the divergence, the
first sum contributes $-\frac\pi2$ to
the return probability. Thus the regularized Laplace transform of the
return probability is
\begin{equation}
-{1\over8}\sqrt{-2z}\cot(L_\perp\sqrt{-2z}),
\end{equation}
if we add all the geometrical prefactors again. This leads to the
equation for the free energy per length of the system with a short
range interaction cited in the main text.

\section{Explicit regularization of the perturbation
series}\label{appexpreg}

If we insert the Laurent expansions of the coefficients $a_i$ into the
perturbation expansion of $u_0(u_R)\over\epsilon\prime$, we get a
regular part of this series to the leading orders in $u_R$ of
\[
a_{1,1}u_R^2+(a_{2,1}+2a_{1,-1}a_{1,2}+2a_{1,0}a_{1,1})u_R^3+O(u_R^4)
\]
As discussed in the main text, the singular parts has a quite simple
structure. All terms of the same divergence order can be combined to
geometric series with polynomial prefactors. The most singular terms
(the first three divergence orders) are
\widetext
\begin{eqnarray*}
\left(a_{1,-1}u_R\over\epsilon^\prime\right)^n\!&\!\Big\{\!&
{1\over a_{1,-1}}+\left[n{a_{1,0}\over a_{1,-1}}+
(n-1){a_{2,-1}\over(a_{1,-1})^2}\right]u_R+\\
&+&\Big[(n+1)a_{1,1}+n{a_{2,0}\over a_{1,-1}}+
n{a_{2,0}\over a_{1,-1}}+
{n(n+1)\over2}{(a_{1,0})^2\over a_{1,-1}}\\
&&\quad +(n^2-n){a_{1,0}a_{2,-1}\over(a_{1,-1})^2}
+{n^2-3n+2\over2}{(a_{2,-1})^2\over(a_{1,-1})^3}\Big]u_R^2
+O(u_R^3)\Big\}
\end{eqnarray*}
\narrowtext
To complete our program, we just have to find the analytic
continuations of series of the form
\begin{equation}
\sum_{n=1}^\infty n^kx^n
\end{equation}
and their limit for $x\to\infty$. This is easy because they all are
derivatives of geometric series. It turns out that the limit for
$x\to\infty$ is $-1$ for $k=0$ and $0$ for all $k>0$. So we get the
contributions of the singular terms in the limit
$\epsilon^\prime\to0$ by just inserting $n=0$ in above expression
and taking the negative value of it. If we do that, we arrive at
\begin{eqnarray}
\lim_{\epsilon^\prime\to0}{u_0(u_R)\over\epsilon^\prime}&=&
-{1\over a_{1,-1}}+{a_{2,-1}\over(a_{1,-1})^2}u_R\\\nonumber
&&-{(a_{2,-1})^2\over(a_{1,-1})^3}u_R^2+
{(a_{2,-1})^3\over(a_{1,-1})^4}u_R^3+O(u_R^4)
\end{eqnarray}
(for the third order coefficient we need one term more in the above
formula for the singular parts, which has been omitted, because it is
easily computed but quite lengthy.)

This is obviously the beginning of a pure geometric series.

If $a_{1,-1}=0$, the regular part of the
series $u_0(u_R)/\epsilon^\prime$ is just
\begin{equation}
a_{1,1}u_R^2+O(u_R^3).
\end{equation}
The singular part consist again of geometric series with polynomial
coefficients and explicitly reads
\widetext
\begin{eqnarray*}
&&{1\over a_{2,-1}u_R}\sum_{n=1}^\infty
\left(a_{2,-1}u_R^2\over\epsilon^\prime\right)^n(1+n a_{1,0}u_R)\\
&+&{u_R\over a_{2,-1}}
\sum_{n=1}^\infty\left(a_{2,-1}u_R^2\over\epsilon^\prime
\right)^n\Bigg[\left({n(n+1)\over2}(a_{1,0})^2+n\,a_{2,0}\right)\\
&&\qquad+\left(n(n+1)a_{1,0}a_{2,0}+(n+1)a_{1,1}a_{2,-1}+
na_{3,0}+{n^3+3n^2+2n\over6}(a_{1,0})^3\right)u_R\Bigg]\\
&+&O(u_R^3)
\end{eqnarray*}
\narrowtext
The limit $\epsilon^\prime\to0$ is again performed by inserting $n=0$
and taking the negative value which reproduces the expected
result~(\ref{u0urwitha1reg}).

\section{First three orders of the partition function for an
arbitrary number of directed polymers}\label{app3ordersN}

During the calculation of the integrands in the series expansion of
the partition function, most of the terms can be strongly simplified
by
\begin{itemize}
\item using the symmetry of the one particle propagator
\item moving parts of the arguments of the one particle
propagator from one argument to the other using the fact that the
propagator depends only on the difference of the arguments
\item translating ${\bf R}^d$-integrations by $kL_\perp$ terms
\item translating ${\bf Z}^d$ summations by $k^\prime$ from other sums
\item combining of sums over ${\bf Z}^d$ and integrals over
$[0,L_\perp]^d$ to integrals over ${\bf R}^d$.
\end{itemize}

With this technique, it can be generally shown that a directed
polymer which is not involved in any of the interactions does not
contribute to the value of a diagram and that a directed polymer that
is involved only in one interaction contributes just factor of
$L_\perp^{-d}$. We will call this lemma~A and represent it
graphically as
\begin{center}
\setlength{\unitlength}{0.006250in}%
\begin{picture}(460,114)(28,701)
\thicklines
\put(299,730){\line( 1, 1){ 40}}
\put(299,744){\line( 1, 1){ 68}}
\put(299,757){\line( 1, 1){ 54}}
\put(299,770){\line( 1, 1){ 41}}
\put(312,730){\line( 1, 1){ 27}}
\put(380,730){\makebox(0.4444,0.6667){.}}
\put(367,730){\line( 1, 1){ 26}}
\put(352,730){\line( 1, 1){ 27.500}}
\put(339,730){\line( 1, 1){ 27.500}}
\put(326,730){\line( 1, 1){ 26.500}}
\put(299,784){\line( 1, 1){ 27.500}}
\put(136,784){\line( 1, 1){ 28}}
\put(149,784){\line( 1, 1){ 28}}
\put(164,784){\line( 1, 1){ 27}}
\put(164,784){\makebox(0.4444,0.6667){.}}
\put(177,784){\line( 1, 1){ 27.500}}
\put(124,784){\line( 5, 6){ 24.016}}
\put(299,798){\line( 1, 1){ 13.500}}
\put( 83,784){\line( 1, 1){ 27.500}}
\put( 96,784){\line( 1, 1){ 28}}
\put( 96,784){\makebox(0.4444,0.6667){.}}
\put(110,784){\line( 1, 1){ 27}}
\put(406,757){\makebox(0.4444,0.6667){.}}
\put(420,784){\line( 1, 1){ 27.500}}
\put(447,812){\makebox(0.4444,0.6667){.}}
\put(420,784){\makebox(0.4444,0.6667){.}}
\put(406,784){\line( 1, 1){ 27.500}}
\put(433,784){\line( 1, 1){ 27.500}}
\put(227,772){\makebox(0,0)[lb]{\smash{$=\!\!L^{-d}$}}}
\put(447,770){\line( 1, 1){ 41.500}}
\put(447,784){\line( 1, 1){ 27}}
\put(392,784){\line( 1, 1){ 28}}
\put(433,730){\line( 1, 1){ 54.500}}
\put(420,730){\line( 1, 1){ 68}}
\put(406,730){\line( 1, 1){ 27}}
\put(392,730){\line( 1, 1){ 27.500}}
\put(380,730){\line( 1, 1){ 26.500}}
\put(447,730){\line( 1, 1){ 40.500}}
\put(380,784){\line( 1, 1){ 27}}
\put(367,784){\line( 5, 6){ 24.016}}
\put(352,784){\line( 1, 1){ 28}}
\put(473,730){\line( 1, 1){ 14.500}}
\put(460,730){\line( 1, 1){ 27.500}}
\put(177,770){\line( 1, 1){ 41}}
\put(394,800){\circle*{2}}
\put(394,793){\circle*{2}}
\put(394,786){\circle*{2}}
\put(394,779){\circle*{2}}
\put(394,807){\circle*{2}}
\put(339,770){\line( 1, 0){108}}
\put( 28,703){\line( 1, 0){189}}
\put(124,703){\circle*{4}}
\put(124,770){\circle*{4}}
\put(394,814){\circle*{2}}
\put(394,772){\circle*{2}}
\put(394,722){\circle*{2}}
\put(299,730){\framebox(189,82){}}
\put(339,757){\framebox(108,27){}}
\put( 28,730){\framebox(189,82){}}
\put( 69,757){\framebox(108,27){}}
\put(394,729){\circle*{2}}
\put(394,764){\circle*{2}}
\put(394,757){\circle*{2}}
\put(394,750){\circle*{2}}
\put(394,743){\circle*{2}}
\put(394,736){\circle*{2}}
\put( 69,770){\makebox(0.4444,0.6667){.}}
\put(136,730){\line( 1, 1){ 27.500}}
\put(124,730){\line( 1, 1){ 26}}
\put(110,730){\line( 1, 1){ 26.500}}
\put( 96,730){\line( 1, 1){ 27.500}}
\put(149,730){\line( 1, 1){ 27.500}}
\put(177,757){\line( 1, 1){ 40.500}}
\put(204,730){\line( 1, 1){ 13.500}}
\put(190,730){\line( 1, 1){ 27}}
\put(177,730){\line( 1, 1){ 40}}
\put(164,730){\line( 1, 1){ 53.500}}
\put( 83,730){\line( 1, 1){ 27}}
\put( 28,770){\line( 1, 1){ 41.500}}
\put( 28,784){\line( 1, 1){ 28}}
\put( 28,798){\line( 1, 1){ 14.500}}
\put(124,770){\line( 0,-1){ 67}}
\put( 69,770){\line( 1, 0){108}}
\put( 28,757){\line( 1, 1){ 55}}
\put( 69,730){\line( 1, 1){ 27}}
\put( 56,730){\line( 1, 1){ 27}}
\put( 43,730){\line( 1, 1){ 26.500}}
\put( 28,730){\line( 1, 1){ 40.500}}
\put( 28,744){\line( 1, 1){ 68}}
\end{picture}
\end{center}
Moreover it is possible to prove lemma~B:
\begin{center}
\setlength{\unitlength}{0.006250in}%
\begin{picture}(444,105)(32,720)
\thicklines
\put(384,754){\line(-1, 0){ 91}}
\put(293,800){\line( 1, 1){ 15}}
\put(293,785){\line( 1, 1){ 30}}
\put(293,770){\line( 1, 1){ 45}}
\put(308,770){\line( 1, 1){ 45}}
\put(323,770){\line( 1, 1){ 45.500}}
\put(338,770){\line( 1, 1){ 45.500}}
\put(384,739){\line(-1, 0){ 91}}
\put(124,739){\line( 1, 1){ 76}}
\put(139,739){\line( 1, 1){ 76}}
\put(155,739){\line( 1, 1){ 60.500}}
\put(170,739){\line( 1, 1){ 45.500}}
\put(185,739){\line( 1, 1){ 30.500}}
\put(200,739){\line( 1, 1){ 15}}
\put(293,815){\line( 0,-1){ 45}}
\put(293,770){\line( 1, 0){ 91}}
\put(384,770){\line( 0,-1){ 31}}
\put(384,739){\line( 1, 0){ 92}}
\put(476,739){\line( 0, 1){ 76}}
\put(476,815){\line(-1, 0){183}}
\put(353,770){\line( 1, 1){ 46}}
\put(461,739){\line( 1, 1){ 15}}
\put(323,754){\line( 0,-1){ 15}}
\put( 62,754){\line( 0,-1){ 30}}
\put( 93,739){\line( 0,-1){ 15}}
\put(223,774){\makebox(0,0)[lb]{\smash{$=\!\!L^{-d}$}}}
\put(446,739){\line( 1, 1){ 30.500}}
\put(369,770){\line( 1, 1){ 45.500}}
\put(384,770){\line( 1, 1){ 45.500}}
\put(384,754){\line( 1, 1){ 61.500}}
\put(384,739){\line( 1, 1){ 76.500}}
\put(400,739){\line( 1, 1){ 76}}
\put(415,739){\line( 1, 1){ 61}}
\put(430,739){\line( 1, 1){ 46}}
\put(124,754){\line( 1, 1){ 61}}
\put(353,746){\circle*{4}}
\put(353,754){\circle*{4}}
\put(353,762){\circle*{4}}
\put(353,770){\circle*{4}}
\put(353,778){\circle*{4}}
\put(353,785){\circle*{4}}
\put(353,793){\circle*{4}}
\put(353,739){\circle*{4}}
\put( 62,724){\circle*{8}}
\put( 93,724){\circle*{8}}
\put( 62,754){\circle*{8}}
\put( 93,739){\circle*{8}}
\put(323,739){\circle*{8}}
\put(323,754){\circle*{8}}
\put(353,731){\circle*{4}}
\put(353,800){\circle*{4}}
\put( 32,770){\line( 1, 1){ 45.500}}
\put( 47,770){\line( 1, 1){ 45.500}}
\put( 62,770){\line( 1, 1){ 45.500}}
\put( 78,770){\line( 1, 1){ 45.500}}
\put( 93,770){\line( 1, 1){ 45.500}}
\put(108,770){\line( 1, 1){ 46}}
\put(124,770){\line( 1, 1){ 45.500}}
\put( 32,785){\line( 1, 1){ 30}}
\put(353,808){\circle*{4}}
\put(353,815){\circle*{4}}
\put(353,823){\circle*{4}}
\put( 32,815){\line( 0,-1){ 45}}
\put( 32,770){\line( 1, 0){ 92}}
\put(124,770){\line( 0,-1){ 31}}
\put(124,739){\line( 1, 0){ 91}}
\put(215,739){\line( 0, 1){ 76}}
\put(215,815){\line(-1, 0){183}}
\put( 32,800){\line( 1, 1){ 15}}
\put( 32,724){\line( 1, 0){183}}
\put(124,754){\line(-1, 0){ 92}}
\put(124,739){\line(-1, 0){ 92}}
\end{picture}
\end{center}
With this preparation, it is easily possible to compute the 1-, 2- and
3-point-function. The 1-point-function has just one diagram with the
prefactor $N(N-1)/2$,
\begin{center}
\setlength{\unitlength}{0.006250in}%
\begin{picture}(80,44)(20,780)
\thicklines
\put( 20,800){\line( 1, 0){ 80}}
\put( 60,820){\line( 0,-1){ 20}}
\put( 60,820){\makebox(0.4444,0.6667){.}}
\put( 20,780){\line( 1, 0){ 80}}
\put( 20,820){\line( 1, 0){ 80}}
\put( 60,820){\circle*{8}}
\put( 60,800){\circle*{8}}
\end{picture}
\end{center}
which has the value $L_\perp^{-d}$ according to lemma~A. Therefore
\begin{equation}
\langle\Phi(t)\rangle_0={N(N-1)\over2}L_\perp^{-d}.
\end{equation}
In the second order there are three types of diagrams with
combinatorial prefactors of 1, 2(N-2) and $(N-2)(N-3)/2$ respectively
(omitting the general prefactor of $N(N-1)/2$).
\begin{center}
\setlength{\unitlength}{0.006250in}%
\begin{picture}(440,68)(20,756)
\thicklines
\put(220,820){\line( 0,-1){ 20}}
\put(180,820){\line( 1, 0){120}}
\put(180,800){\line( 1, 0){120}}
\put(180,780){\line( 1, 0){120}}
\put(180,760){\line( 1, 0){120}}
\put( 60,820){\line( 0,-1){ 20}}
\put(340,760){\line( 1, 0){120}}
\put(100,820){\line( 0,-1){ 20}}
\put(420,780){\line( 0,-1){ 20}}
\put(260,800){\line( 0,-1){ 20}}
\put(380,820){\line( 0,-1){ 20}}
\put(340,820){\line( 1, 0){120}}
\put(340,800){\line( 1, 0){120}}
\put(340,780){\line( 1, 0){120}}
\put( 20,820){\line( 1, 0){120}}
\put(100,820){\circle*{8}}
\put(380,800){\circle*{8}}
\put(380,820){\circle*{8}}
\put(220,800){\circle*{8}}
\put(220,820){\circle*{8}}
\put( 60,800){\circle*{8}}
\put( 60,820){\circle*{8}}
\put(100,800){\circle*{8}}
\put( 20,800){\line( 1, 0){120}}
\put( 20,780){\line( 1, 0){120}}
\put( 20,760){\line( 1, 0){120}}
\put(260,780){\circle*{8}}
\put(420,780){\circle*{8}}
\put(420,760){\circle*{8}}
\put(260,800){\circle*{8}}
\end{picture}
\end{center}
The last two are reduced by lemma~A to $L_\perp^{-2d}$, whereas the
first one has the value
\[
L_\perp^{-d}\int\limits_{{\bf R}^d}{\rm d}^dr
\sum\limits_{k\in{\bf Z}^d}
G_{Ls_1}(0,r+kL_\perp)G_{Ls_1}(0,r)\equiv R^p(s_1)
\]
Integrating over $r$ results in
\[
R^p(s)=\left({1\over\sqrt{4\pi Ls}}\sum_{k\in{\bf Z}}e^{-{k^2\over4s}}
\right)^d=\left(\frac1{\sqrt{L}}\sum_{k\in{\bf Z}}
e^{-4\pi k^2s}\right)^d,
\]
where the second equation comes from the fact that the sum is the
value of a theta function at zero~\cite{bellman}.

Combining everything, we get in the second order
equation~(\ref{secondorderN}).
The third order consists of 16 different diagrams. All of them
but one can be evaluated by applying lemmas A and B and in the
end we get equation~(\ref{thirdorderN}).


\end{document}